\documentclass[12pt]{iopart}

\expandafter\let\csname equation*\endcsname\relax
  \expandafter\let\csname endequation*\endcsname\relax
\usepackage[usenames,dvipsnames]{color} 
\usepackage{graphicx,amssymb,amsmath}
\usepackage{doi}
\DeclareMathSymbol{\theta}{\mathalpha}{letters}{"23}
\DeclareMathSymbol{\phi}{\mathalpha}{letters}{"27}
\renewcommand{\emph}{\textit}
\usepackage{hyperref}
\usepackage{CJK}
\usepackage{subcaption}

\newcommand{\carb}{$^{13}$C }
\newcommand{\Nit}{$^{14}$N }
\newcommand{\Tone}{$T_1^e$}
\newcommand{\Ttwostar}{$T_2^{*n}$}
\newcommand{\Ttwo}{$T_2^n$}

\newcommand{\ket}[1]{\left\vert{#1}\right\rangle}
\newcommand{\bra}[1]{\left\langle{#1}\right\vert}

\begin{document}
\begin{CJK*}{UTF8}{fs} 

\title{Protecting solid-state spins from a strongly coupled environment}

\author{Mo Chen \CJKfamily{gbsn}(陈墨)$^{1,2}$, Won Kyu Calvin Sun$^{1,3}$, Kasturi Saha$^{1,4}$, Jean-Christophe Jaskula$^{1}$, and Paola Cappellaro$^{1,2}$}
\address{$^1$
Research Laboratory of Electronics, Massachusetts Institute of Technology, Cambridge, Massachusetts 02139, USA
}
\address{$^2$
Department of Mechanical Engineering, Massachusetts Institute of Technology, Cambridge, Massachusetts 02139, USA
}

\address{$^3$
Department of Nuclear Science and Engineering, Massachusetts Institute of Technology, Cambridge, Massachusetts 02139, USA
}

\address{$^4$
Department of Electrical Engineering, Indian Institute of Technology Bombay, Mumbai 400 076, India
}

\ead{pcappell@mit.edu}
\date{\today}
\begin{abstract}
Quantum memories are critical for solid-state quantum computing devices and a good quantum memory requires both long storage time and fast read/write operations. A promising system is the Nitrogen-Vacancy (NV) center in diamond, where the NV electronic spin serves as the computing qubit and a nearby nuclear spin as the memory qubit. Previous works used remote, weakly coupled \carb nuclear spins, trading  read/write speed for long storage time.
Here we focus instead on the intrinsic strongly coupled \Nit nuclear spin. We  first quantitatively understand its decoherence mechanism, identifying as its source the electronic spin that acts as a quantum fluctuator. We then propose a scheme to protect the quantum memory from the fluctuating noise by applying dynamical decoupling on the environment itself. We demonstrate a factor of 3 enhancement of the storage time in a proof-of-principle experiment, showing the potential for a quantum memory that combines fast operation with long coherence time. 
\end{abstract}
\maketitle
\end{CJK*}

\section{Introduction}\label{sec:intro}
Quantum technologies, especially those based on solid-state systems such as superconducting qubits~\cite{Devoret13}, Nitrogen-Vacancy (NV)  centers in diamond~\cite{Doherty13,Degen17}, and dopant spins in silicon~\cite{Zwanenburg13}, have seen significant progress over the past few decades. Qubits embedded in solid-state systems are advantageous because of their compatibility with existing semiconductor fabrication techniques that can offer avenues for scalability. The drawback, however, is their intrinsic noisy environment due to strong couplings to their solid-state host. The fluctuating environment renders qubits fragile, leaving demonstrations of even small scale quantum computing devices (20-50 qubits) challenging~\cite{Bernien17}. 

While further improvements can come from more carefully engineering the qubit systems to remove   undesired noise sources and reduce the number of decoherence channels, achieving fault tolerance will still require some form of quantum error correction (QEC).
Recent developments include both theoretical proposals for more powerful QEC protocols~\cite{Terhal15} and experimental attempts at correcting or detecting quantum errors~\cite{Waldherr14,Taminiau14,Cramer16,Linke17}.
Despite these advances, we have rarely seen experiments yielding better error rate of the error-corrected qubit than the best single qubit in the same system~\cite{Ofek16}. This is because  the recovery operation needed for QEC has so far introduced more error than it corrected.  
A simpler QEC strategy, avoiding measurement and recovery operations, 
is to decouple qubits from the environment using dynamical decoupling (DD).  
This technique, going back to NMR's spin echo~\cite{Hahn50}, 
 enjoys great success thanks to its ease of implementation. In addition,  it is compatible with many quantum information processing protocols~\cite{Khodjasteh09l,Cappellaro09,Zhang14c} and can be concatenated  with active QEC~\cite{Paz-Silva13,Byrd02,Boulant02}. Still, DD has traditionally been applied to refocus slow-varying, weakly coupled environments that can often be modeled as classical bath~\cite{Biercuk11}, while its usefulness to decouple from strongly interacting quantum environments is less clear~\cite{Ban09}.

Here, we explore the effectiveness of DD to increase the coherence time of 
a spin qubit in the presence of a strongly interacting quantum fluctuator. We first introduce and test a quantitive model of the decoherence process of a qubit (the \Nit nuclear spin of a Nitrogen-Vacancy center) subject to random telegraph noise (RTN) arising from the fluctuation of either a spin-$1/2$ or spin-$1$. 
Then, based on the model, we find the requirement on the DD control sequence that achieves qubit protection from the RTN.
It turns out that, due to the slow control on nuclear spin compared to the hyperfine interaction $A$, any DD sequence applied to the nuclear spin would not meet  the requirement  and still yield the same coherence time,  \Ttwostar. However, we find that by modulating the noise source  itself we can   efficiently refocus its effects: control on  the NV electronic  spin  is fast enough to satisfy the DD requirement, and can extend the qubit coherence time beyond the limit imposed by the fluctuator noise. Finally, we realize a proof-of-principle demonstration of these ideas, by protecting the \Nit ~nuclear spin from RTN of a short-lived effective electronic spin-$1/2$. 

\section{Fluctuator Model \& Experiment}

 Random telegraph noise (RTN), often responsible for $1/f$ noise, is ubiquitous in solid-state nanodevices~\cite{Paladino14} and is often the main source of decoherence for quantum dots~\cite{Bermeister14} and most notably for superconducting qubits~\cite{Burnett14,Bermeister14,deGraaf17}.
Here we focus on another exemplary system, nuclear spin qubits in the presence of a fluctuating electronic spin. Specifically, we consider a
 quantum register consisting of the electronic spin-$1$ of NV (in the following, we will refer to this simply as NV), its native \Nit ~nuclear spin-$1$, and possibly a few close-by \carb ~nuclear spins. 
With this system, researchers have demonstrated quantum information storage~\cite{Maurer12,Dolde14},  quantum error correction~\cite{Waldherr14,Taminiau14,Cramer16}, quantum feedback control~\cite{Hirose16} and high-sensitivity magnetometry~\cite{Aslam17,Zaiser16}, taking advantage of the long dephasing time \Ttwostar ~of the nuclear spin, which is usually one to two orders or magnitude longer than that of the NV. Long though it is, \Ttwostar ~is limited by the NV relaxation time \Tone ~($\sim$ few ms)~\cite{Maurer12,Zaiser16,Shim13}. 
Random NV flips due to \Tone ~process result in a 3-level RTN; the  
nuclear spin picks up a random phase from the RTN and decoheres. 
As explained below, when the hyperfine interaction strength  is larger than the \Tone ~flip rate, \Ttwostar ~is strictly limited by \Tone. To extend \Ttwostar ~beyond this limit, previous efforts have focused 
on weakly coupled nuclear spins, employing motional narrowing~\cite{Maurer12,Jiang08}, or  decoherence-free subspace \cite{Reiserer16,WangFan17}. 
On the other hand, strongly coupled nuclear spins are favorable because they provide fast~\cite{Chen15,Sangtawesin15,Shim13}, and direct control. Here, we look into the regime of strongly coupled nuclear spins, where previous methods do not work well. In particular, we work with the native \Nit ~nuclear spin, because it is ubiquitous and has proven useful in the NV-\Nit ~quantum register~\cite{Dolde14,Neumann10b,Aslam17,Zaiser16,Hirose16}.  For this system, the previous approaches do not work well, nor does implementing a simple spin echo~\cite{Shim13}. After gaining a deeper insight into the fluctuator model, we will show how to overcome this challenge. 

\subsection{Spin-Fluctuator modeled as a random walker}
\label{sec:modelnoDD}

We consider a system of two spins interacting via an hyperfine coupling $A$ that we describe semi-classically using a spin-fluctuator model~\cite{Bergli07}. 
\begin{figure}[thb]
 \centering
    \begin{subfigure}[t]{0.5\textwidth}
        \includegraphics[width=0.8\textwidth]{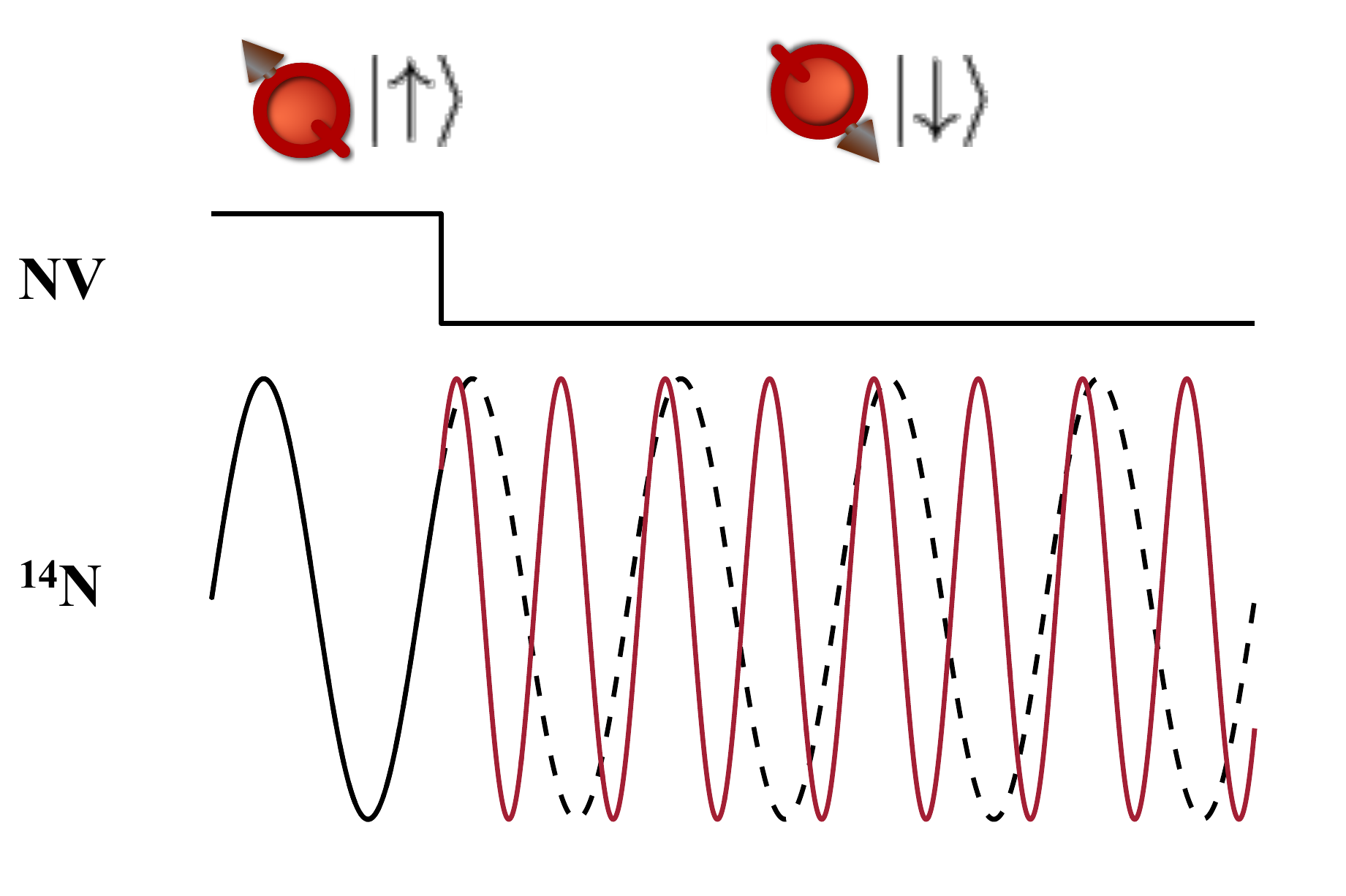}
    \end{subfigure}
\caption{Qubit decoherence under  random telegraph noise. The fluctuator (NV) randomly flips between its two eigenstates (here from $\ket{\uparrow}$ to $\ket{\downarrow}$) changing the rate at which the qubit (\Nit) accumulates a phase. For a representative RTN trace, we show that in the absence of a fluctuator jump, the qubit population would continue to oscillate at the same rate (dashed line), while after a fluctuator jump, the oscillation rate accelerates (solid red line). 
As the jump timing is random, the observed average dynamics is decoherent.}\label{fig:RTNconcept}
\end{figure}

We model the intrinsic \Nit nuclear spin $I$ of the NV center as a random walker, whose phase evolves subject to the state of its neighboring electronic spin $S$ that acts as a strongly coupled fluctuator and generates RTN (Fig.~\ref{fig:RTNconcept}). In such analogy, the velocity $v$ of the random walker is linked to the phase accumulation rate of the qubit, and is set by the hyperfine interaction, $v=m_S A$.
The fluctuator flips at a rate $\gamma$ between any two $m_S$ states due to  spin-lattice relaxation,  inducing a  change in the value of the  hyperfine coupling,  $m_S A$,  thus  at each such event the walker's velocity $v$ also changes. In between fluctuator jumps, the walker covers a distance $\phi$, representing the accumulated phase between qubit states. 

For a 2-level fluctuator (2LF), the random walker has only two possible directions of motion, left or right. The probability $\hat p_r$($\hat p_l$) of reaching a position $\phi$ from the right(left) is therefore the sum of probabilities of keeping going right(left) from $\phi-vdt$ or turning right(left) (with probability $\gamma$) from $\phi+ v dt$. The random walker's motion can then be described by a system of differential equations~\cite{Bergli07}:
\begin{equation}
	\left\{\begin{array}{rcl}
		\partial_t \hat p_l(\phi, t)&=&-\gamma[\hat p_l(\phi, t)-\hat p_r(\phi, t)]+v\partial_\phi \hat p_l(\phi, t)\\
\partial_t \hat p_r&=&-\gamma(\hat p_r-\hat p_l)-v\partial_\phi \hat p_r.
	\end{array}\right.
\end{equation}
 The spin coherence is given by the average accumulated phase, $\langle e^{i\phi}(t)\rangle=\int \hat P(\phi, t) e^{i\phi} d\phi$, where $\hat P(\phi, t)=\hat p_l+\hat p_r$  is the total probability of reaching $\phi$ at time $t$. Note that this is the Fourier transform of $\hat P$,  $P(k,t)$, evaluated at $k=-1$. We can thus more compactly write the relevant equations of motion as $\partial_t \vec{P}(k,t)=M_k\vec{P}(k,t)$, that is

\begin{equation}
\frac{\partial}{\partial t}
\begin{pmatrix}
P\\p
\end{pmatrix}
=
\begin{pmatrix}
0&-ikv\\-ikv&-2\gamma
\end{pmatrix}
\begin{pmatrix}
P\\p
\end{pmatrix}, 
\end{equation}
where $p(k,t)=p_r(k,t)-p_l(k,t)$.

We can similarly describe the spin dynamics in the presence of a three-level fluctuator (3LF), where one of the levels corresponds to a ``rest'' state $v=0$ (no phase accumulation), with corresponding probability $p_0$:

\begin{equation}
\frac{\partial}{\partial t}
\begin{pmatrix}
P\\p^+\\p
\end{pmatrix}
=
\begin{pmatrix}
0&0&-ikv\\2\gamma&-3\gamma&-ikv\\0&-ikv&-3\gamma
\end{pmatrix}
\begin{pmatrix}
P\\p^+\\p
\end{pmatrix}, 
\end{equation}
where $p^+=p_l+p_r$ and here $P=p_l+p_r+p_0$. The spin decay is then given by $\langle e^{i\phi}(t) \rangle=e^{M_{-1}t} P(-1, 0)$ and is characterized by a typical timescale $T_2^{*n}$. Note that since we consider the fluctuator to be either a spin-1/2 or a spin-1, we  have $v=A/2$($A$) and  $\gamma=1/2$\Tone($1/3$\Tone) for the 2LF(3LF). 

In the strong fluctuator regime that we focus on, intuitively  a single fluctuator jump  is enough to totally decohere the nuclear spin. Then, provided $v/\gamma\gg1$, the spin decay rates doe not depend anymore on $v$ but only on the jump rate. Indeed, we find  $T_2^{*n}=2T_1^e(1.5T_1^e)$ for the 2LF(3LF) (See Appendix.~\ref{sec:strongfluctuatorregime}). 
This strict limit on \Ttwostar ~for the strong fluctuator is in sharp contrast to the weak fluctuator case, where \Ttwostar ~increases as the hyperfine interaction strength decreases (See Appendix.~\ref{sec:weak3LF}).

\subsection{Experimental results}\label{sec:natural}
To experimentally test the spin-fluctuator model, we measure and compare \Tone ~of the  Nitrogen-Vacancy center electronic spin and \Ttwostar ~of its native \Nit nuclear spin.
All experiments are performed using a home-built confocal microscope, with single NV centers in an electronic grade diamond sample (Element 6, \Nit concentration $[^{14}N]<5$~ppb, natural abundance of \carb). We work at a magnetic field of $424$G, close to the excited state level anti-crossing, to polarize the \Nit nuclear spin~\cite{Jacques09}.  
A $1.5$mW laser of $2\mu$s duration  polarizes the hybrid NV-\Nit system into $\ket{m_S=0,m_I=+1}$ with high fidelity. Microwave (MW) and radiofrequency (RF) pulses are delivered through a $25\mu$m wide copper wire to have precise control of the NV and \Nit spin states.
 
For \Tone~measurement, a laser pulse first initializes the system into $\ket{0,+1}$. Then we apply a strong MW pulse ($t_\pi =44$~ns) to prepare it to the desired state $\ket{-1,+1}$. The NV is free to fluctuate due to \Tone ~process before we measure the remaining population in $\ket{-1,+1}$ obtaining the signal  $S_{-1}^{-1}$ where the  sub(super)script refers to the initial (final) electronic spin state. We also measure the population in the state $\ket{0,+1}$ obtaining the signal $S_{-1}^{0}$. \Tone~is extracted to be $4.3\pm 0.3$~ms by fitting to the difference of the two measurements $S_{-1}^{-1}-S_{-1}^{0}$ (here and throughout the paper, uncertainty in all fitted values are $95\%$ confidence interval).

For \Ttwostar~measurement, we implement a nuclear Ramsey sequence  in the electronic $m_s=-1$ manifold, where the larger nuclear spin energy splitting (due to the hyperfine coupling) allows faster driving. The system is first prepared to $\ket{-1,+1}$, as described above, before being coherently driven to a nuclear superposition state $(\ket{-1,+1} + \ket{-1,0})/\sqrt{2}$ using on-resonant RF field. After a free evolution period, we convert nuclear spin coherence to populations with a second RF $\pi/2$ pulse, $\alpha \ket{-1,+1} + \beta\ket{-1,0}$. Finally, the nuclear spin is read out by mapping its state to the NV electronic spin, using a selective MW pulse ($t_\pi=1.1\mu$s). This pulse  creates the entangled state $\alpha \ket{0,+1} + \beta\ket{-1,0}$ between the NV and \Nit, allowing optical readout of the nuclear spin state population $\lvert\alpha\rvert^2, \lvert\beta\rvert^2$ with high SNR. 

Isolating the bare contribution of the nuclear spin dephasing in the fluorescence signal decay is nonetheless not straightforward as the consequences of NV random flips are three-fold. First, they induce the nuclear spin dephasing that we aim at measuring. Second, they modify the NV state, leading the second RF $\pi/2$ pulse to be off-resonance and preventing the transfer of the nuclear coherence into  population difference. Finally, they also induces errors in the mapping between nuclear and electronic spin states as the NV state is not fully polarized anymore. 

\begin{figure*}
 \centering
    \begin{subfigure}[t]{0.48\textwidth}
        \caption{}\label{fig:naturalT1}
        \includegraphics[width=0.8\textwidth]{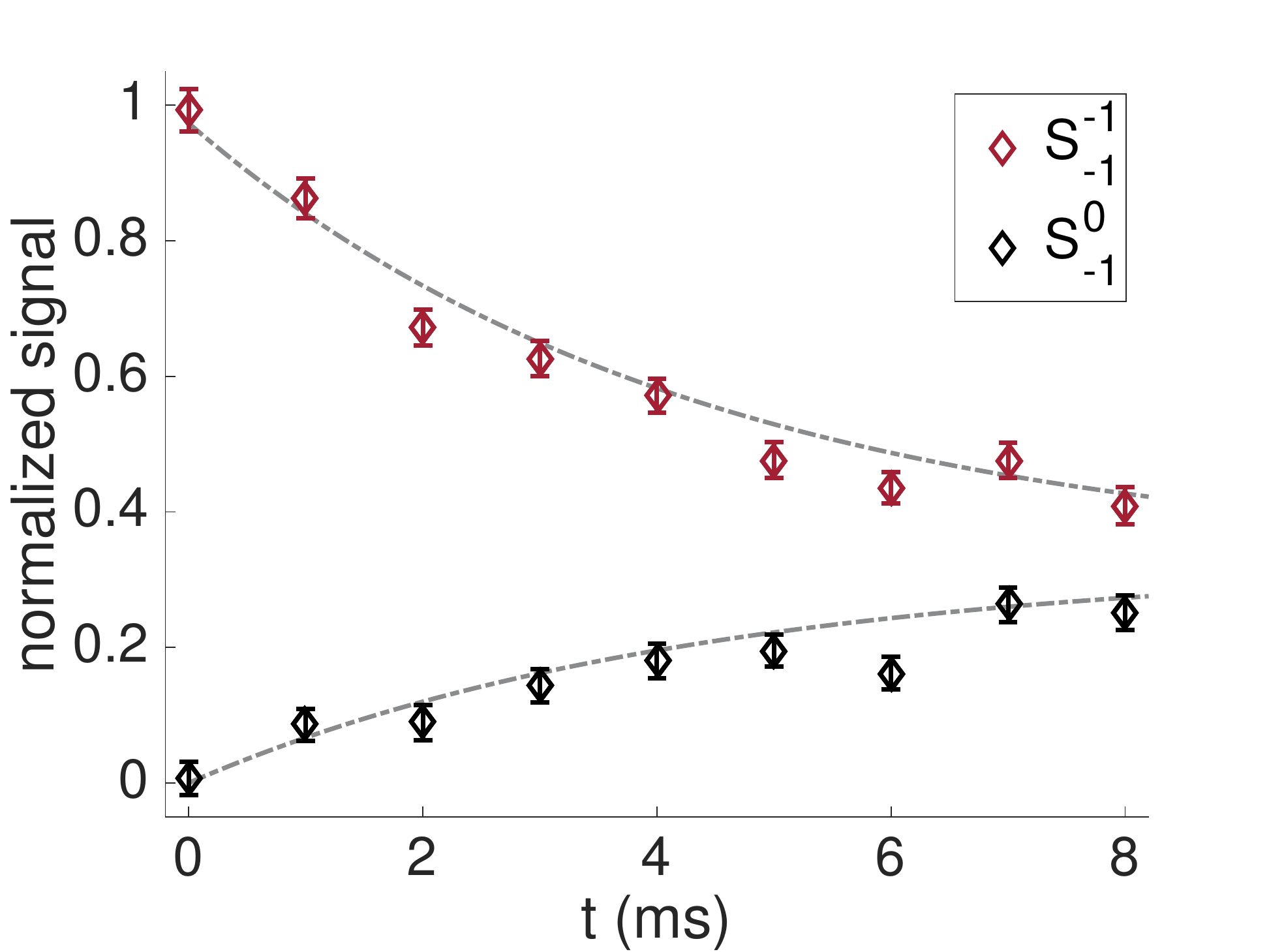}
    \end{subfigure}
  ~
    \begin{subfigure}[t]{0.48\textwidth}
        \caption{}\label{fig:naturalT2star}
        \includegraphics[width=0.8\textwidth]{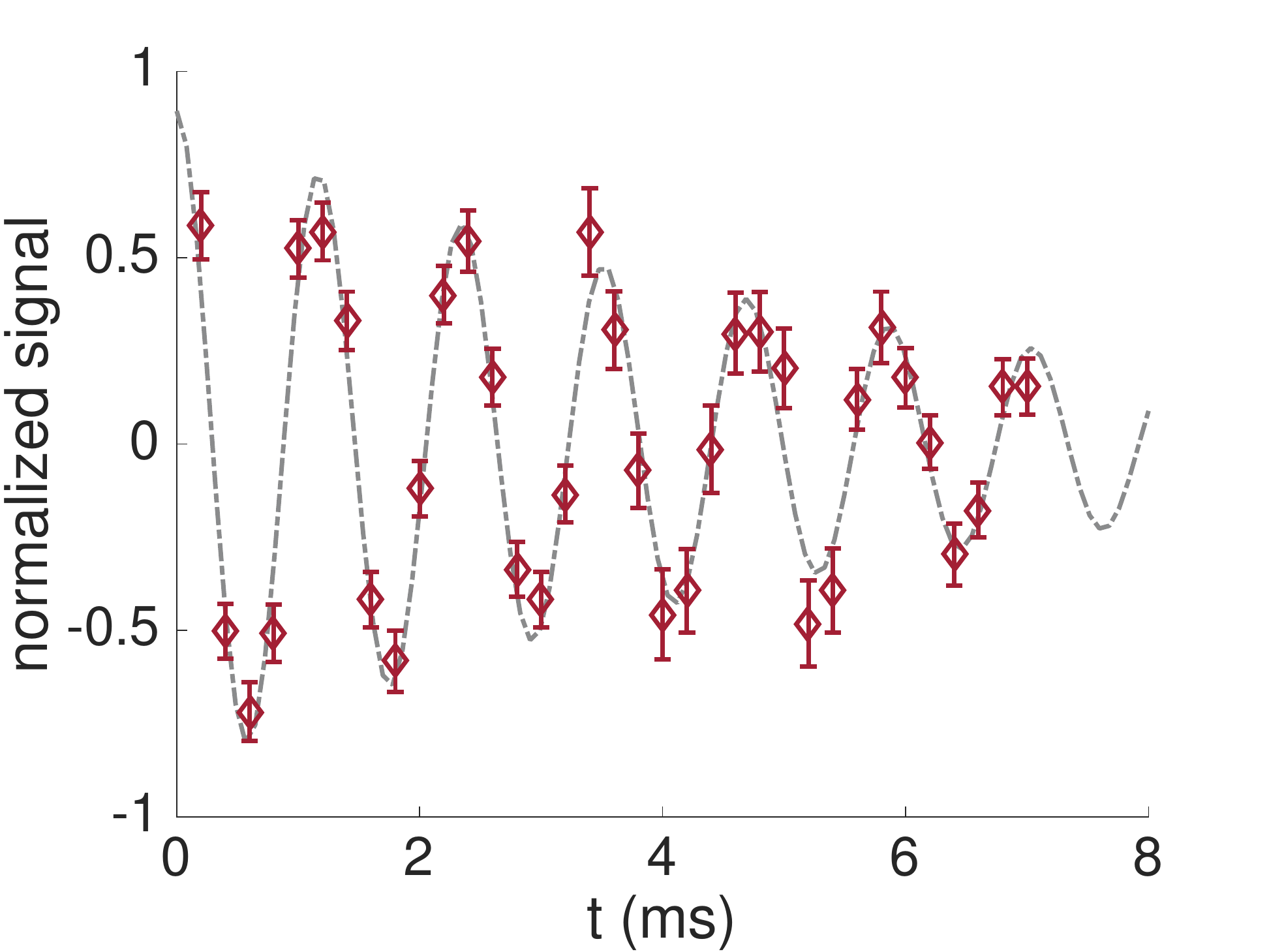}
    \end{subfigure}
\caption{(a) Natural relaxation  decay of a single NV population and (b) cohernce decay of its native \Nit under a Ramsey sequence. The dashed lines are fits to the expected dynamics, yielding \Tone$=4.3(3)$~ms and \Ttwostar$=5.6(1.7)$~ms. These values satisfy $T_2^{*n}=1.5 T_1^e$, as predicted by the spin fluctuator model. All error bars in the figure are one standard error of the mean (SEM).}
\end{figure*}

Fortunately, in the strong fluctuator regime, one flip of the fluctuator is enough to decohere our qubit, which allows us to neglect the two last errors.
The bare contribution of the nuclear spin dephasing can be isolated by recording the signals obtained from nuclear Ramsey sequence with 1) no phase difference between the two RF $\pi/2$ pulses and 2) a $\pi$ phase shift. The last two effects that cause imperfect readout of the nuclear spin have an equivalent contribution in both Ramsey sequences, just creating a common error that is suppressed when subtracting the two Ramsey signals (Appendix.~\ref{sec:diffmess}). 
We can then measure a dephasing time \Ttwostar $=5.6\pm1.7$~ms as shown in Fig.~\ref{fig:naturalT2star}. This measurement is consistent with our prediction from spin-fluctuator model with 3 levels, $T_2^{*n}=1.5 T_1^e$.

\section{Dynamical Decoupling in the strong coupling regime}
\subsection{Theory}\label{sec:modelDD}
To protect the nuclear spin qubit from RTN generated by the NV, we resort to dynamical decoupling (DD). Usual DD schemes are highly effective in protecting qubits from noise provided the pulses are applied at a higher repetition rate than the typical correlation time of the noise. Unfortunately, because of the Markovian nature of RTN, this condition does not apply here. Instead of being set by the fluctuator rate $\gamma$, in order for DD to be effective the $\pi$-pulse separation time $\tau$ must satisfy
\begin{equation}\label{eq:effective_condition}
A\cdot\tau\lesssim1.
\end{equation} 

When applying DD following the well-known CPMG  sequence~\cite{Meiboom58} with time between pulses $\tau$, the spin coherence after $N$ pulses is given by 
\begin{equation}\label{eq:probability}
\langle e^{i\phi} (N\tau) \rangle = \{[e^{M_{-1}\tau/2}\cdot U_\pi \cdot e^{M_{-1}\tau/2}]^N \vec{P}\}_1,
\end{equation}
where $U_\pi$ is the $\pi$-pulse operator in the $\vec{P}$ basis. For a 2LF this is diag$(1,-1)$, while  it is diag$(1,1,-1)$ for a 3LF.
As discussed below, it might be possible, and even more convenient, to apply $\pi$-pulses on the fluctuator instead of the qubit. Indeed, the desired effect is to invert the sign of the coupling between the two systems. For a 3LF, this can be achieved by driving the double quantum (DQ) transition $\ket{+1}\leftrightarrow\ket{-1}$. However,  for a 3LF there is some freedom on the type of pulses applied. In addition to driving the DQ transition,  one can also drive one of the  single quantum (SQ) transitions, $\ket{0}\leftrightarrow\ket{\pm1}$, resulting in $$U_\pi=
\begin{pmatrix}1&0&0\\1&-1/2&\mp1/2\\1&\mp3/2&1/2\end{pmatrix}.$$ 

Since the qubit decay under DD is not necessarily purely exponential~[Appendix.~\ref{sec:T23LF}], we define an effective coherence time $T_2^n(\tau)$ through:
\begin{equation}\label{eq:effT2}
\langle e^{i\phi}(N\tau=T_2^n(\tau)) \rangle=1/e
\end{equation}
The dependence of $T_2^n$ on the DD interval $\tau$ is shown in Fig.~\ref{fig:DDsimulation}. As expected, smaller $\tau$'s are better at decoupling the qubit from RTN  and at extending \Ttwo. Interestingly, the behavior for the 2LF and 3LF is different. For 2LF, DD leaves the decay approximately exponential (see Appendix.~\ref{sec:T22LF}), with a decay rate
\begin{equation}\label{eq:2levelT2}
1/T_2^n=\gamma-\frac{1}{\tau} \ln\left[\frac{\gamma \sin({W}\tau) + \sqrt{v^2 - \gamma^2 \cos^2({W}\tau)}}{{W}}\right],
\end{equation}
where $W=\sqrt{v^2-\gamma^2}$.
For a 3LF, however, the coherence decay is not exponential (See Appendix.~\ref{sec:T23LF}). 
Still, we see that, by using DQ pulses to refocus the fluctuator (or applying pulses directly to the qubit), one could in principle fully decouple the qubit from RTN noise when $\tau\to0$, {until nuclear-nuclear dipolar interactions become the dominant noise source~\cite{Maurer12}}. 
With SQ drive, however, one only protects the qubit from RTN half of the time, therefore the  decay rate can at most be reduced to  half of its value without DD (Fig.~\ref{fig:DDsimulation}).

\subsection{Experimental results}
\label{sec:eng}
We now  apply these ideas  to  protect the nuclear spin from the NV RTN and extend its \Ttwo ~beyond the limit of \Tone. 

Due to the strong hyperfine coupling $A=2.16$~MHz between the electronic  and nuclear spin of the NV, the condition (\ref{eq:effective_condition}) requires $\tau\lesssim1\mu$s, which is not feasible given the slow control of the nuclear spin qubit (with typical $\pi$-pulse times $50\mu$s). However, as the hyperfine interaction is symmetric with respect to the state of both spins, applying $\pi$-pulses on either the qubit or the fluctuator modulates the hyperfine interaction sign and will lead to an effectively weaker averaged hyperfine coupling and thus a slower rate at which nuclear states acquire a random phase. 
It is consequently possible to take benefit from the electronic driving strength that are typically a few tens to a few hundreds of MHz~\cite{Fuchs09,DeLange10,Scheuer14}, yielding $\pi$-pulses fast enough to meet the requirement of Eq.~(\ref{eq:effective_condition}). 

Another challenge in the experiment is due to the nuclear spin readout, which is indirectly obtained by measuring the NV spin. As mentioned in Sec.~\ref{sec:natural}, as the NV center state is unknown at the end of the evolution due to \Tone ~processes, the nuclear spin mapping from coherence to population states and its readout via the NV electronic spin might fail. This problem is exacerbated when  DD is applied, as we expect some qubit coherence to be stored in all NV manifolds. Thus, the differential measurement scheme applied above no longer provides an accurate picture of the nuclear spin coherence decay. In particular, it is no longer possible to fully measure gains in \Ttwo ~beyond \Ttwostar ~(see Appendix.~\ref{sec:diffmess}). 

\begin{figure}[t]
    \centering
     \begin{subfigure}[t]{0.48\textwidth}
        \caption{}\label{fig:fakeT1sequence}
        \includegraphics[width=0.95\textwidth]{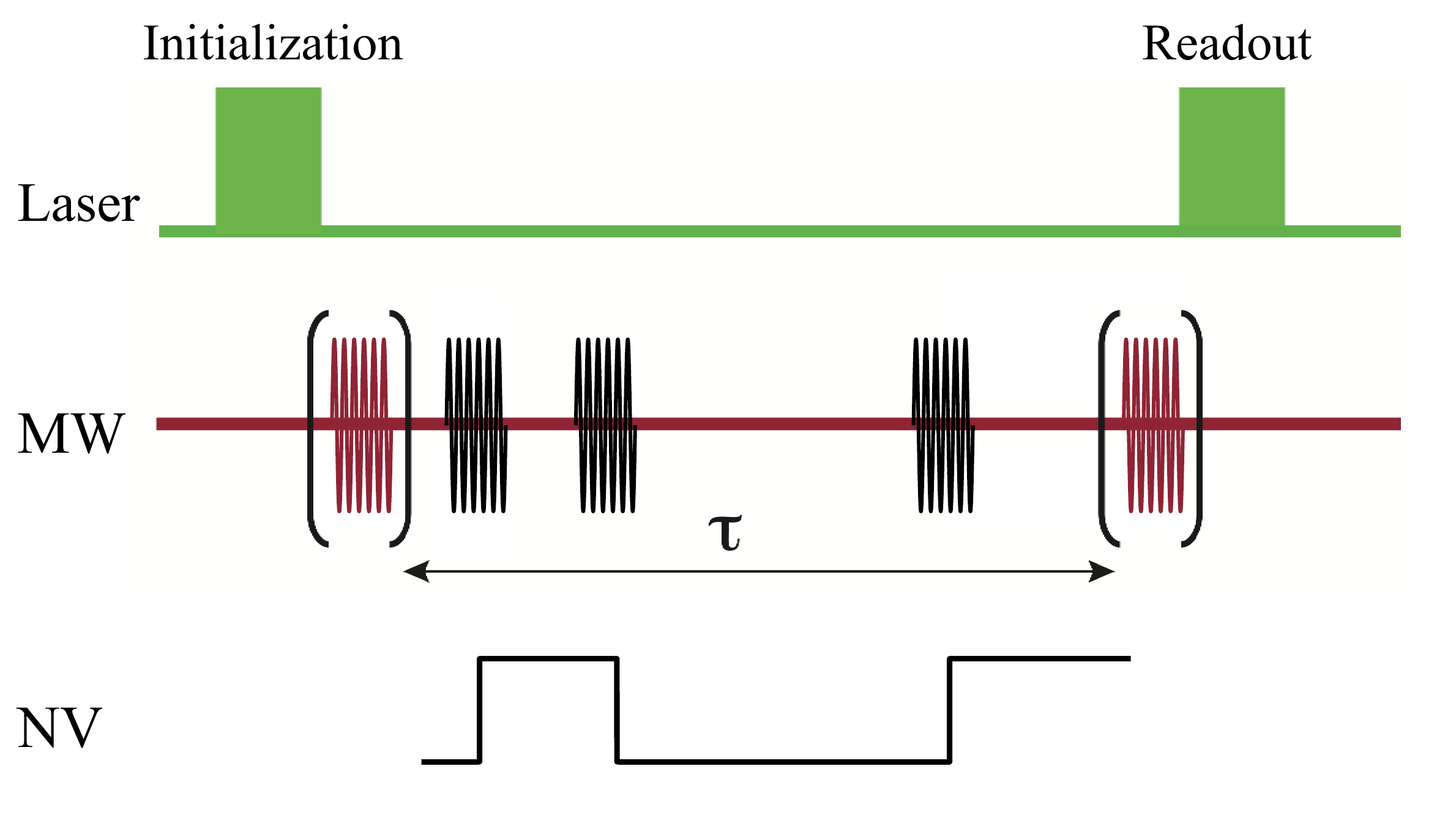}
    \end{subfigure}
~
    \begin{subfigure}[t]{0.48\textwidth}
        \caption{}\label{fig:fakeT1}
        \includegraphics[width=0.8\textwidth]{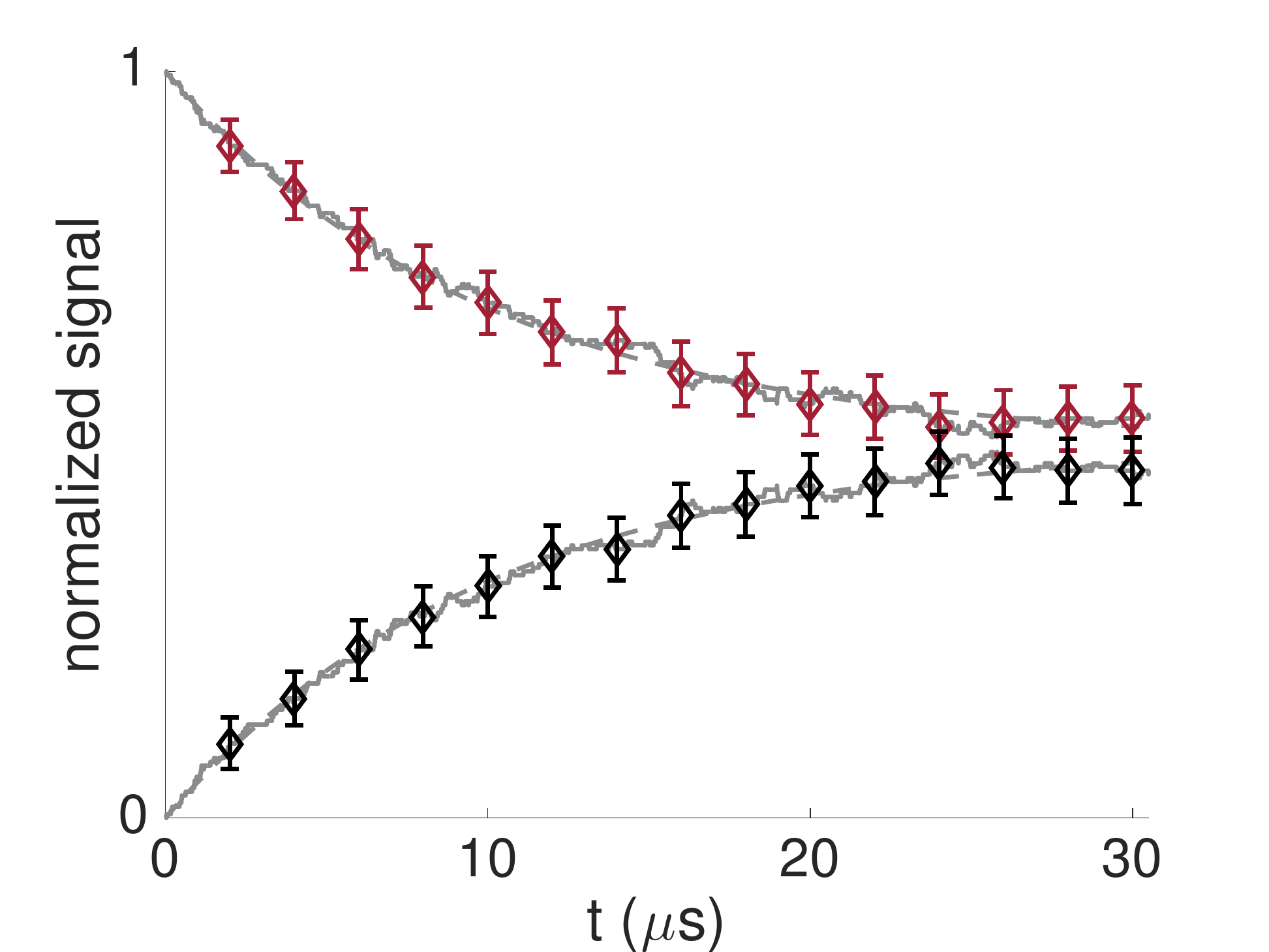}
    \end{subfigure}%
    \caption{(a) Pulse sequence for one engineered \Tone ~trace {with three fluctuator jumps}. The red MW $\pi$-pulses prepare and readout the desired NV state. The black MW $\pi$-pulses mimic engineered \Tone ~flips. (b) Decay of a single NV under engineered \Tone ~relaxation noise, simulated by 200 traces of engineered \Tone ~flips as described in the main text.  Red and black diamonds: $S_{-1}^{-1}$ and $S_{-1}^{0}$ experimental decays. Solid gray line: simulation of \Tone ~using the same traces. Gray dashed line: fit to an exponential decay, giving \Tone~$=10.0(4)\mu$s. Error bars in the figure are one SEM.}
\end{figure}

To remove this undesired effect and more precisely verify the  protection of the nuclear spin afforded by DD, we engineer  a  short-lived 2LF, decoupling its evolution (and final state) from the state needed for the correct readout of the nuclear spin. The engineered noise also allows shorter experiments, further avoiding slow external experimental drifts that could hide the gains in coherence time. In addition, engineering a 2LF instead of 3LF eliminates the need of dual frequency driving.
The artificial 2LF is engineered by applying fast, on-resonant MW pulses to flip the NV electronic spin state between $\ket{0}$ and $\ket{-1}$ (engineered \Tone ~flip) at random times following a Poisson distribution. Figure~\ref{fig:fakeT1sequence} displays one of the $200$ engineered \Tone ~traces that once averaged simulate an exponential \Tone ~decay process. 
We set the flipping constant of the artificial 2LF in order to obtain a relaxation time \Tone $=10\mu$s. 
This time scale is much longer than the $\pi$-pulse length $t_p=44$ns, and is two orders smaller than that of the natural \Tone, guaranteeing the third level $\ket{+1}$ of the NV center is not involved in the dynamics and we indeed have an effective 2-level fluctuator. 

\begin{figure}[t]
 \centering
    \begin{subfigure}[t]{0.48\textwidth}
        \caption{}\label{fig:DDsimulation}
        \includegraphics[width=0.8\textwidth]{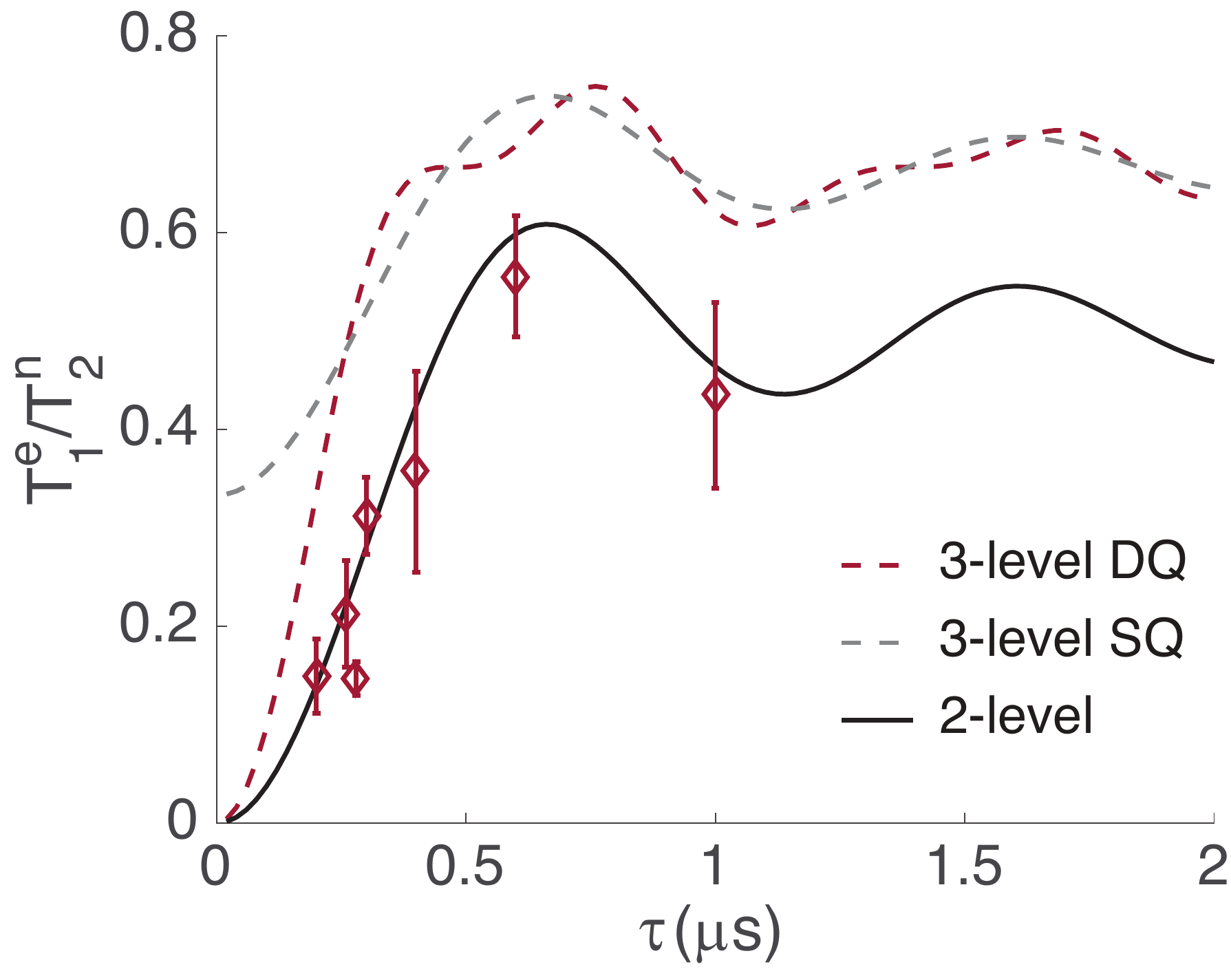}
    \end{subfigure}
    \begin{subfigure}[t]{0.48\textwidth}
        \caption{}\label{fig:fakeT2}
        \includegraphics[width=0.8\textwidth]{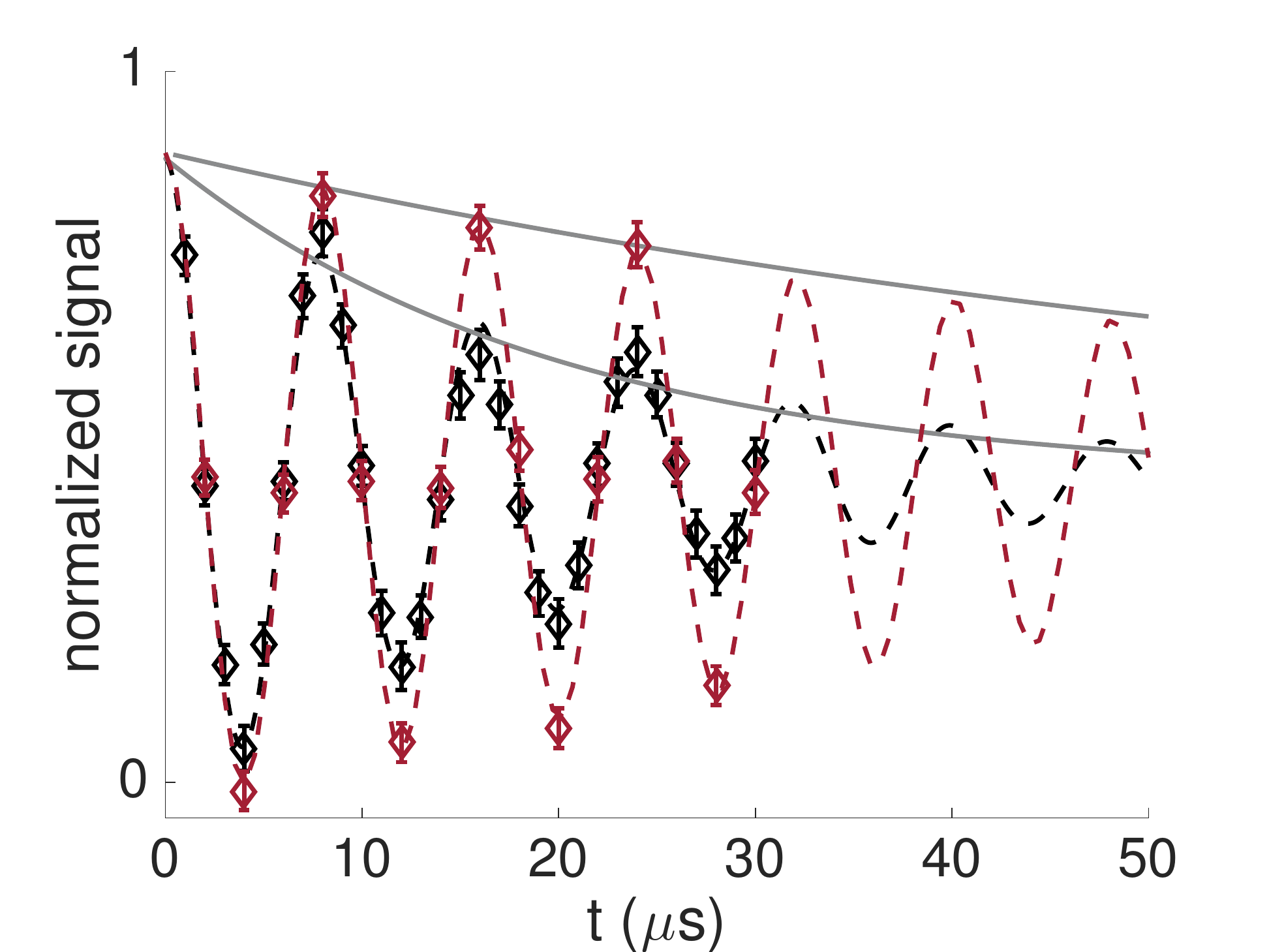}
    \end{subfigure}
 \caption{(a) Effective coherence time $T_2^n$ for 2LF (theory: black solid line; experiment: red diamond), and 3LF for double quantum drive (theory, red dashed line) and single quantum drive (theory, gray dashed line).  The experimental results (for  $\tau=200,260,280,300,400,600$, and $1000$~ns) agree well with the theoretical prediction of the 2LF, including the somewhat counterintuitive result for $\tau=600$ns which indeed gives \Ttwo$<$\Ttwostar. Error bars are $95\%$ confidence interval; (b) Corresponding \Ttwostar ~(experiment, black diamond) and \Ttwo ~(experiment, red diamond, $\tau=200$ns DD interval) decay. Dashed lines are fits to an oscillating exponential decay. Gray solid lines are the theoretical \Ttwostar ~and \Ttwo ~decays calculated according to  the spin-fluctuator model (amplitude renormalized). Error bars are one SEM (see Appendix~\ref{sec:appendixexp}).}
\end{figure}

Figure.~\ref{fig:fakeT1} shows measurement of the engineered \Tone ~decay, matching very well with the simulation of the applied 200 engineered \Tone ~traces. An exponential fit gives a decay time of $10.0\pm 0.4\mu$s, in good agreement with our $10\mu$s design. We then use the same engineered \Tone ~traces to perform a nuclear Ramsey experiment and measure the resulting coherence time by fitting data to an oscillating exponential curve, obtaining an engineered \Ttwostar $=22\pm4\mu$s $=2$\Tone ~(Figure~\ref{fig:fakeT2}), as expected from the 2LF theory. In order for the nuclear spin state readout to be accurate, if the NV ends up in $\ket{0}$ due to the engineered flips we apply an extra $\pi$-pulse immediately before the readout process, which brings it back to $\ket{-1}$.  

Finally, in addition to the engineered \Tone ~traces, we apply the Knill-Dynamical Decoupling (KDD) sequence~\cite{Souza11KDD} with an interval of $\tau=200$ns on the NV to decouple \Nit from the RTN. We choose KDD instead of CPMG because it is robust against pulse errors (In the following experiments, we use DD to refer to KDD pulses). 
We apply $\sim100$~$\pi$ pulses to measure the nuclear spin coherence decay, which according to Eq.~\ref{eq:2levelT2}, is expected to follow an exponential behavior. 
In Fig.~\ref{fig:fakeT2} we compare the experimental (and theoretical) decays with and without DD, clearly showing the improvement achieved by applying a decoupling scheme, proving the successful protection of nuclear spin from its RTN environment. This is confirmed by the measured \Ttwo ~value, extracted from an oscillating exponential fit to be \Ttwo $=67\pm 17\mu$s, (Fig.~\ref{fig:fakeT2}), clearly exceeding \Ttwostar, and matching well with theoretical prediction of $71\mu$s. We repeat this experiment with different DD intervals $\tau$, 
to compare the trend in \Ttwo ~with our theory. The results, shown  in Fig.~\ref{fig:DDsimulation}, are in quantitative agreement with the predicted behavior, including the somewhat counterintuitive result for $\tau=600$ns, where the applied DD \textit{accelerates} decoherence, giving \Ttwo$<$\Ttwostar. 

While we demonstrated that dynamical decoupling can be effective in increasing the coherence time affected by a strong random fluctuator only for engineered noise, we remark that our experimental results show that it would be possible to refocus the natural noise as well. Indeed, using a direct readout of the nuclear spin \cite{Robledo11,Neumann10b} or with a different protocol to map its state onto the electronic spin, it would be possible to avoid seeing direct effects of the NV \Tone ~on the measured nuclear spin coherence. We further performed DD experiments to show that it is practically feasible to implement the necessary number of pulses for decoupling (see Appendix~\ref{sec:naturalwDD}) without introducing additional noise due to pulse errors. 

\section{Conclusion}
Protecting a qubit strongly coupled to a fluctuating quantum environment is often a challenging task. Here we studied an exemplary system comprising the electronic and \Nit nuclear spins associated with the NV center in diamond. While the nuclear spin can act as a long-lived qubit (or memory), the electronic spin, which is necessary for initialization and readout, is also the main source of noise for the qubit. 
We theoretically analyzed the decoherence mechanism of the nuclear spin qubit and introduced a simple model in terms of a random fluctuator to describe its decoherence.
Measurements on the fluctuator and \Nit qubit are consistent with our spin-fluctuator model and show the limit on the qubit coherence. Based on this model, we proposed a method to decouple and protect the nuclear spin from its environment, and demonstrated a factor of $3$ increase in coherence time in a proof-of-principle experiment.

Our results pave the way to using strongly coupled nuclear spins, including the ubiquitous native Nitrogen of the  NV center, for demanding experiments requiring long quantum memory times, complementing existing techniques applicable only to weakly coupled nuclear spins~\cite{Maurer12,Reiserer16,WangFan17}. In addition, the proposed technique based on DD is compatible with many quantum information processing protocols~\cite{Khodjasteh09l,Cappellaro09,Zhang14c}, allowing the full functionality of a quantum register, where the electronic spin performs local operations while the quantum memory is protected. This is in contrast to other protocols where the electronic spin is inaccessible during protection of nuclear spins~\cite{Cohen17,Maurer12,Jiang08}. Proposals concatenating DD with active QEC~\cite{Paz-Silva13,Byrd02,Boulant02} also makes it potentially a first layer of protection before applying QEC, enabling scaling-up with less overhead. Finally, the proposed control technique is also applicable to other solid-state systems, for example, superconducting qubits, where single or ensembles of fluctuators are believed to be the major noise source~\cite{Bergli07,simmonds04,Paladino14}.

\section*{Acknowledgements}
We thank Pai Peng for helpful discussion. 
This work was supported in part by the NSF grants EECS1702716 and 1641064, by the ARO MURI W911NF-15-1-0548 and by ONR N00014-14-1-0804.

\section{Appendix 1: Experimental Methods}\label{sec:appendixexp}

\subsection{Differential Measurement of Nuclear Spin Coherence}\label{sec:diffmess}
In our experiments we do not have a direct measurement of the nuclear spin qubit, which would reveal its coherence time. A common strategy to overcome this limitation is to initialize the NV and apply a CNOT gate, flipping the NV state conditional on the nuclear spin state. This effectively maps the state information from the nuclear to the electronic spin. In the nuclear Ramsey experiment in Sec.~\ref{sec:natural}, we create the entangled state $\alpha\ket{0,+1}+\beta\ket{-1,0}$ with this protocol. The NV is then optically read out, giving the same probability distribution $\lvert\alpha\rvert^2, \lvert\beta\rvert^2$ as if directly measuring the nuclear spin. However, when we measure the nuclear spin coherence, the NV undergoes \Tone ~flips, potentially introducing errors in this readout process. First, if the NV final state is different from the nominal one, the second RF pulse operates in the incorrect NV manifold and is thus off-resonance, failing to transfer the nuclear coherence into state populations. Second, the mapping between nuclear and electronic spin states might fail as the NV state is not fully polarized anymore. 

\begin{figure}
    \centering
    \begin{subfigure}[t]{0.48\textwidth}
        \caption{}\label{fig:T2starreadout}
        \includegraphics[width=0.8\textwidth]{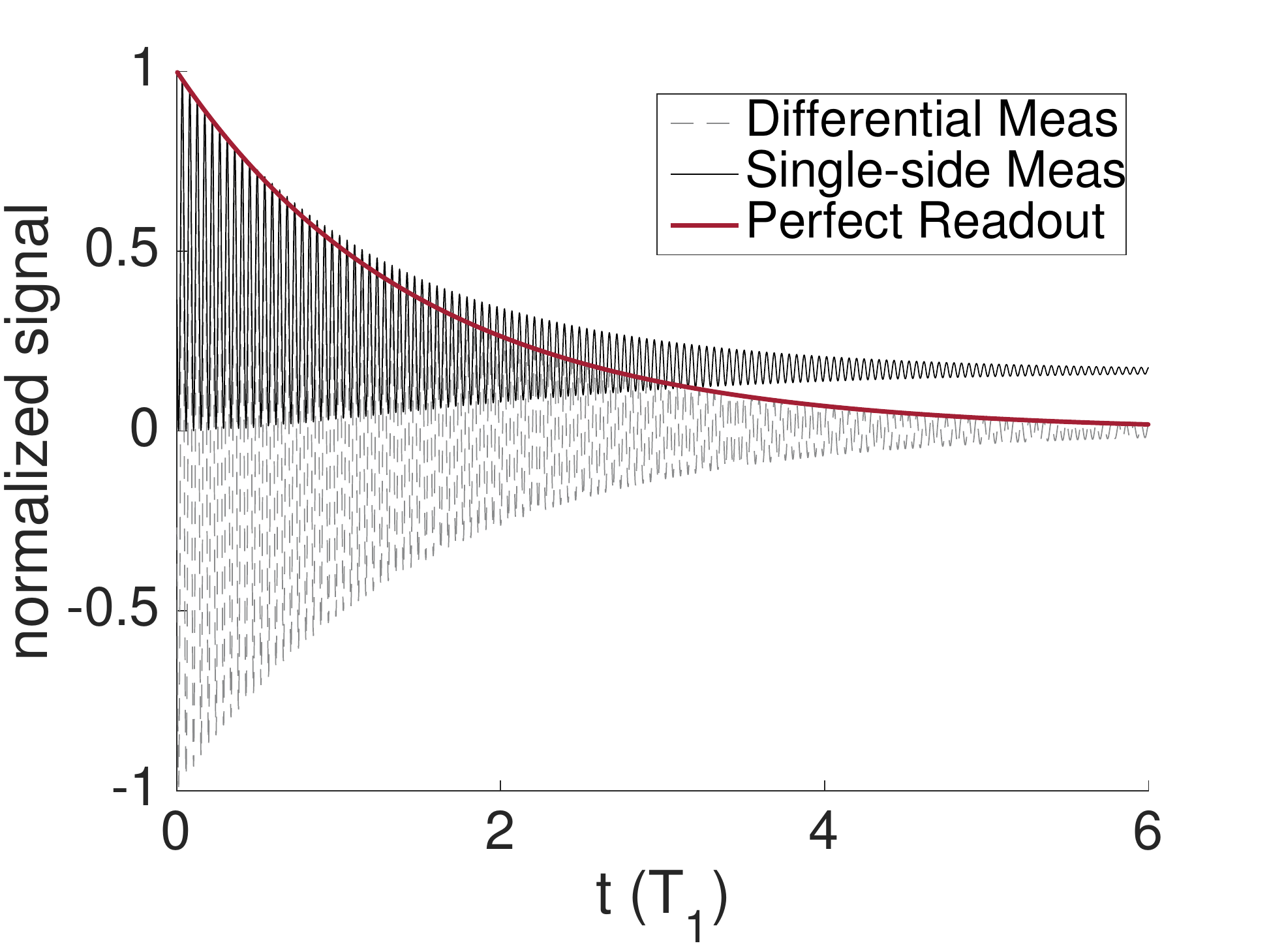}
    \end{subfigure}
    ~
    \begin{subfigure}[t]{0.48\textwidth}
        \caption{}\label{fig:T2readout_diff}
        \includegraphics[width=0.8\textwidth]{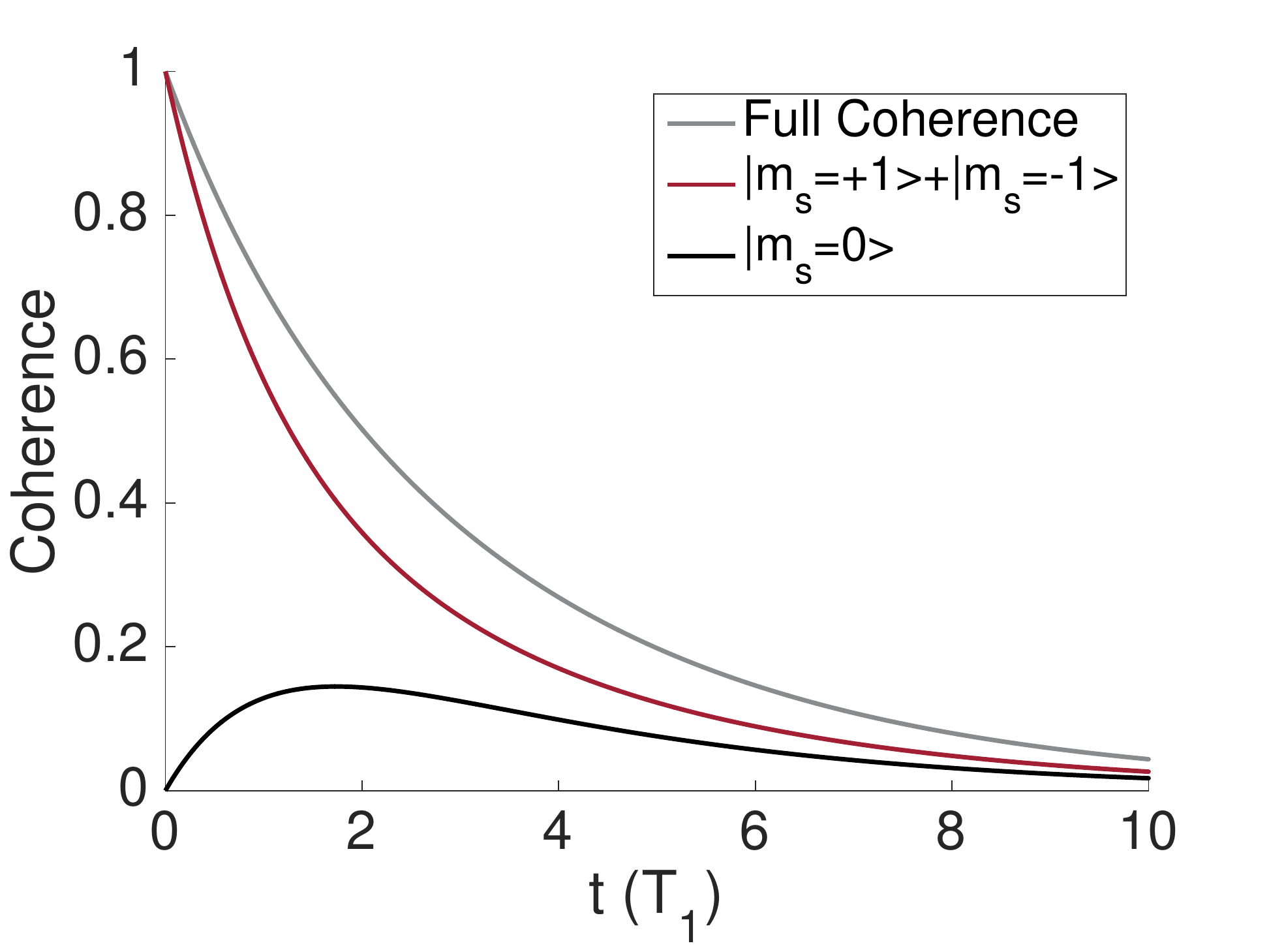}
    \end{subfigure}
    \caption{(a) Simulation of differential measurement for \Ttwostar. Red solid line assumes perfect readout of the nuclear spin coherence. Black solid line is one set of the differential data. The asymmetric shape and non-zero asymptotic value indicate the presence of a common mode signal not related to nuclear coherence. Gray dashed line shows differential measurement, revealing \Ttwostar. (b) Nuclear coherence stored in different NV manifolds when we apply DQ DD. Gray solid line is the full coherence. Red solid line is the coherence stored in $\ket{m_s=\pm 1}$ and black solid line  in $\ket{m_s}=0$. As coherence is stored in all manifolds, the differential measurement is no longer effective in removing common mode noise from the NV \Tone ~process.}
\end{figure}

To account for these errors, we perform a differential readout to obtain \Ttwostar: the first measurement is the regular nuclear Ramsey experiment, measured by applying a CNOT on the NV, subject to possible readout failure; for the second measurement, we add a $\pi$ phase shift on the second RF $\pi/2$ pulse in the nuclear Ramsey sequence, and then apply the same readout. The difference of the two measurements yields the expected \Ttwostar ~decay, as shown in Fig.~\ref{fig:T2starreadout}. The intuitive explanation is that since one jump of the strong fluctuator totally decoheres the qubit, any nuclear spin coherence is preserved only when the NV stays in the original state. No coherence is left when the final state is different. Therefore, measurements (including unsuccessful ones) already contain all  the information about the nuclear spin coherence. The purpose of the differential readout is to remove the two errors mentioned above, which are not related to the nuclear spin coherence decay. Therefore, one could even choose a different measurement as the second data set, as long as it contains the same common mode error. One such choice is to apply a CNOT gate conditional on the nuclear spin being in the $\ket{m_I=0}$ state (instead of $\ket{m_I=+1}$). We chose to apply a $\pi$ phase shift on the nuclear Ramsey because when taking the difference of the two measurements this also doubles the signal amplitude.
When the qubit is protected against RTN by the DD control, this differential measurement is no longer effective because some nuclear spin coherence is stored in all NV manifolds (Fig.~\ref{fig:T2readout_diff}). 
We emphasize that this is not a fundamental limit. Single-shot readout, either using on-resonant laser at cryogenic temperature~\cite{Robledo11} or using intermediate-high magnetic field at room-temperature~\cite{Neumann10b} solves this issue and has been experimentally demonstrated.

\begin{figure}[b]
    \centering
    \begin{subfigure}[t]{0.48\textwidth}
        \caption{}\label{fig:singleshotfixedt}
        \includegraphics[width=0.8\textwidth]{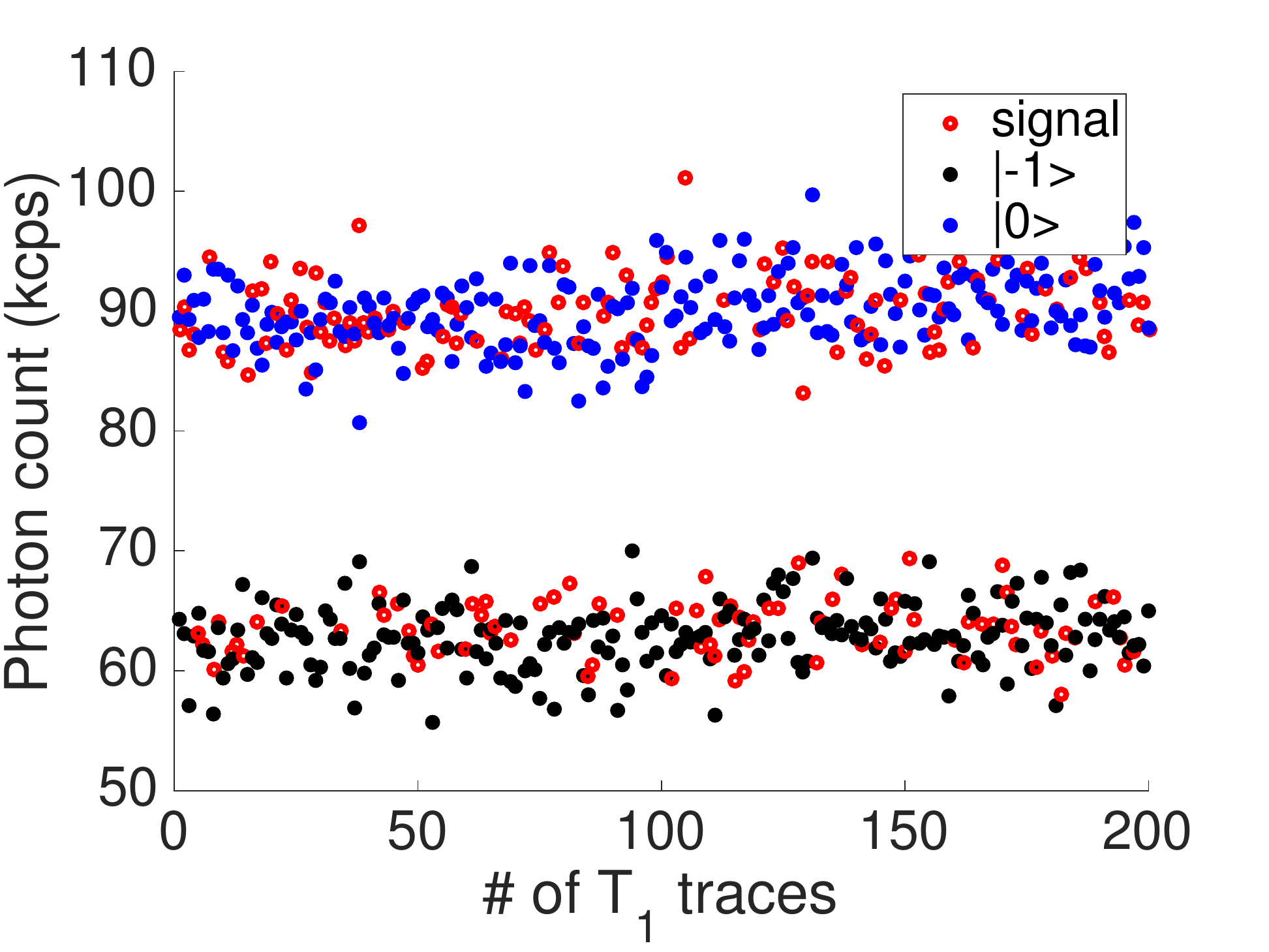}
    \end{subfigure}
    ~ 
    \begin{subfigure}[t]{0.48\textwidth}
        \caption{}\label{fig:fakeT1trace}
        \includegraphics[width=0.8\textwidth]{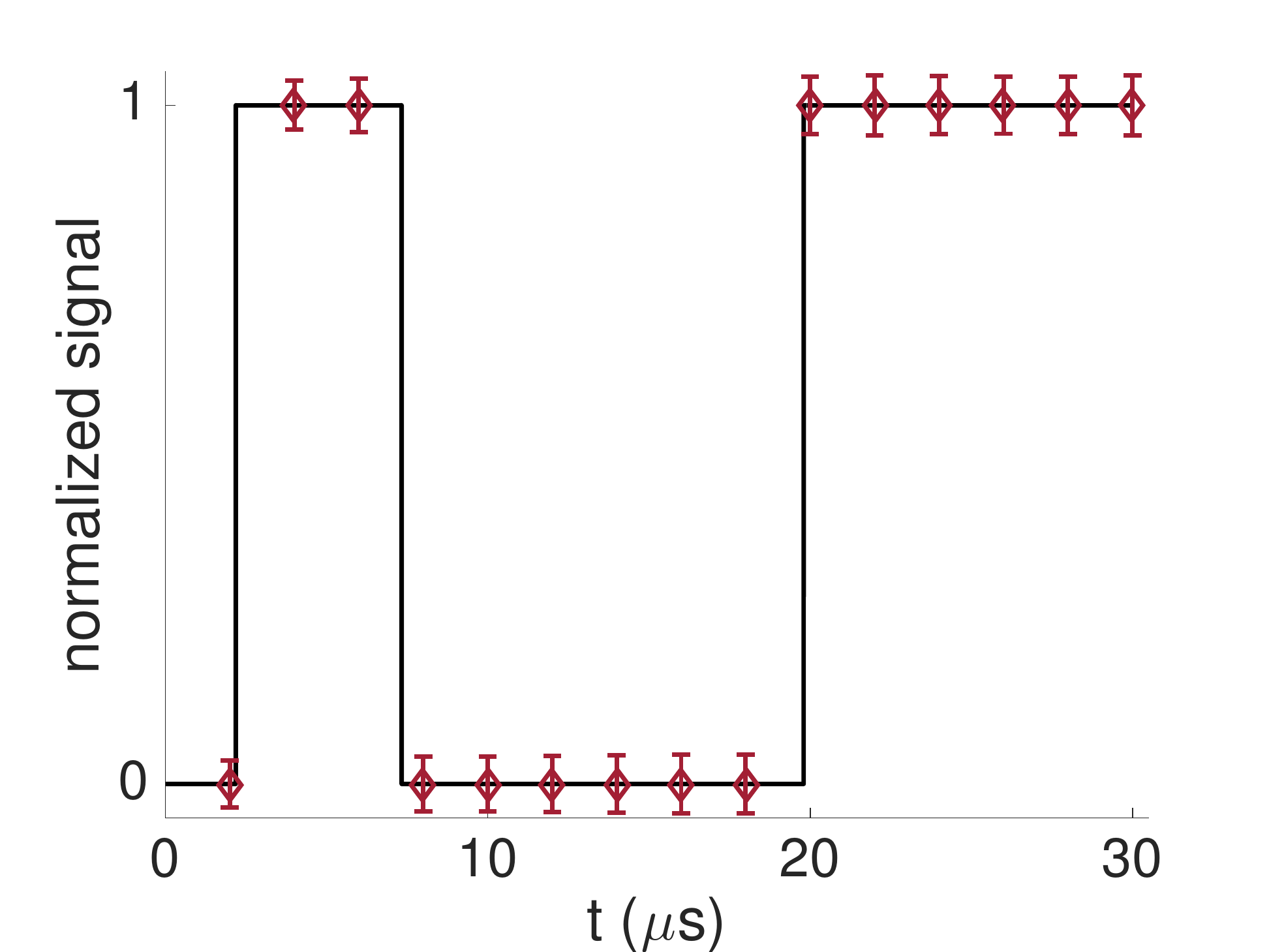}
    \end{subfigure}
    \caption{Engineered \Tone ~measurement. (a) We compare the measured NV fluorescence at a fixed time, t$=16\mu$s,  to the ``bright'' and ``dark''  reference lines, given by the population states $\ket{0}$, $\ket{-1}$, for all $200$ different traces. We clearly see that the final state can be read out with high fidelity. (b) One of the engineered \Tone ~trace of theory (black solid line) and experiment (red diamond). Error bars are one SEM.}
\end{figure}
\subsection{Data Analysis for Engineered \Tone}
In the engineered \Tone ~experiment, we consider $N_T=200$ predetermined traces. Each trace is repeated $4\times10^4$ times in order to build enough statistics to determine the final population state of the NV. In Fig.~\ref{fig:singleshotfixedt}, we plot the average signal at t$=16\mu$s for each one of the $200$ \Tone ~traces, clearly showing that the final state of each trace can be reliably determined to be  either $\ket{0}$ or $\ket{-1}$. As another demonstration, in Fig.~\ref{fig:fakeT1trace} we plot out the $\# 47$ engineered \Tone ~trace in black solid line, and the experimental result in red diamonds. 
The error bars in all engineered \Tone ~experiments are then the standard deviation resulting from the $200\times4\times10^4$ acquired data for each time point, divided by $\sqrt{N_T}$, which corresponds to the usual standard error of the mean used for the other experimental results.

\subsection{Engineered \Tone~with DD}
\begin{figure*}
    \centering
    \begin{subfigure}[t]{0.48\textwidth}
        \caption{}\label{fig:fakeT1DD}
        \includegraphics[width=0.8\textwidth]{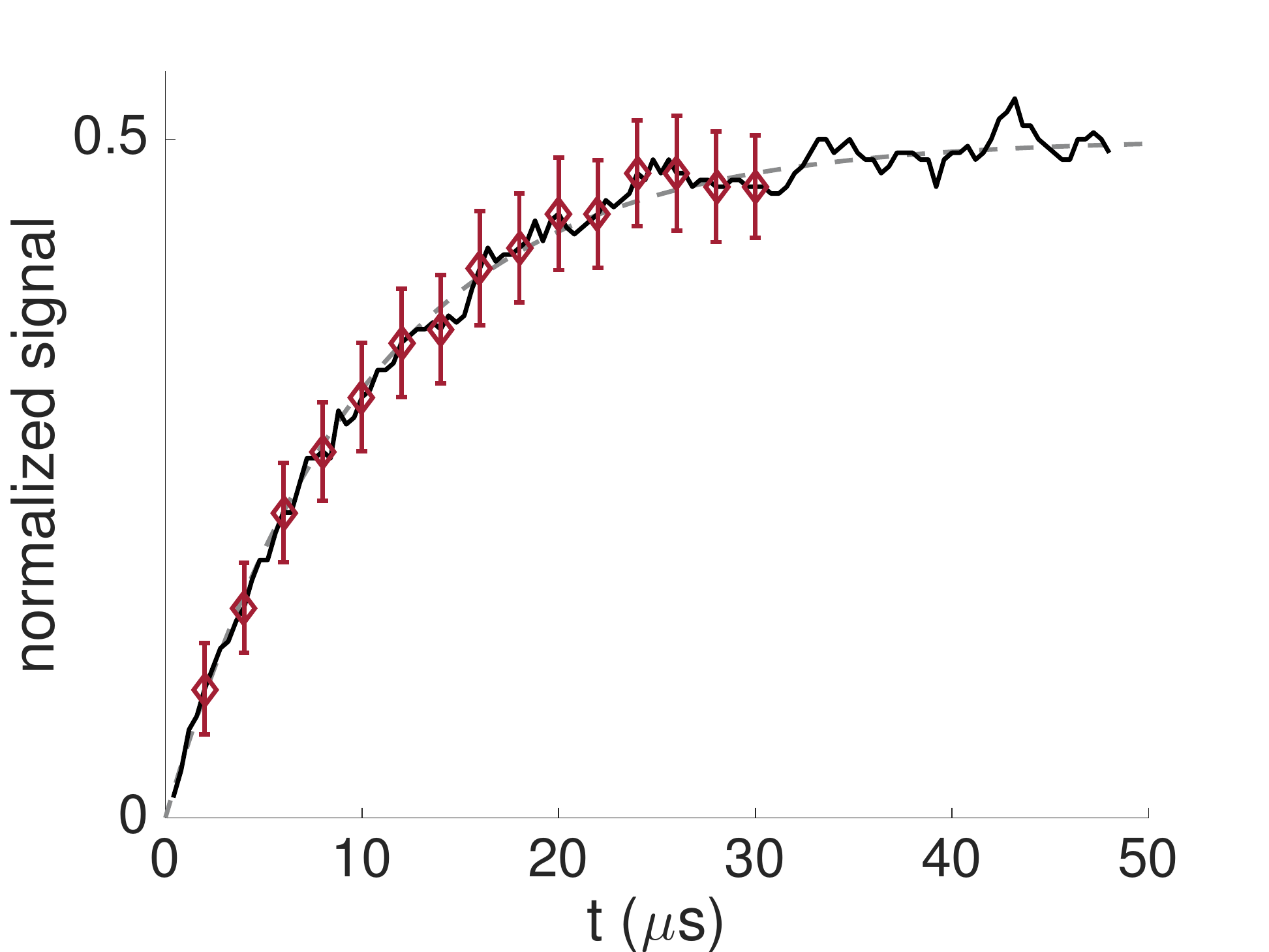}
    \end{subfigure}
    ~ 
    \begin{subfigure}[t]{0.48\textwidth}
        \caption{}\label{fig:naturalT1DD}
        \includegraphics[width=0.8\textwidth]{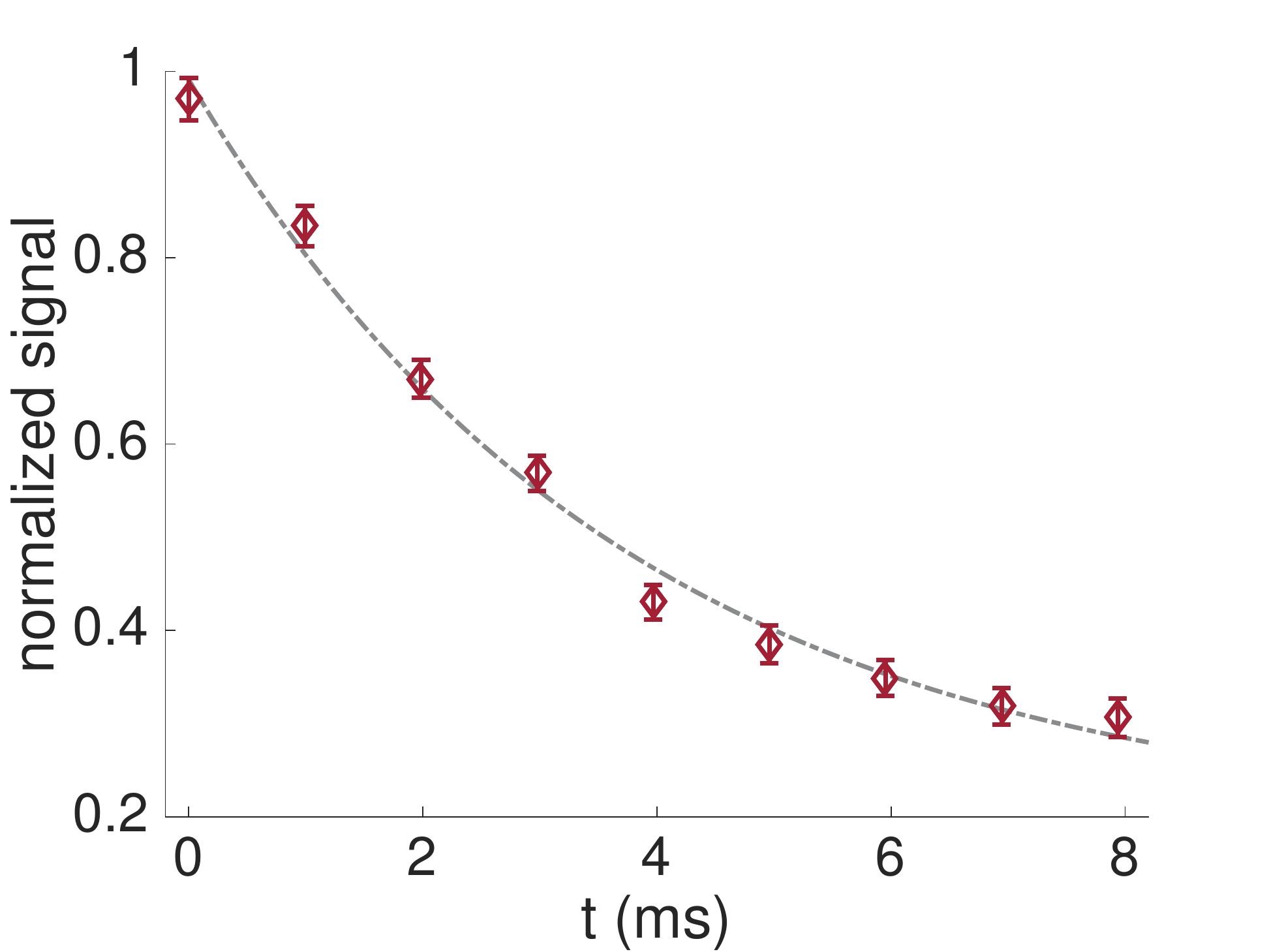}
    \end{subfigure}
    \caption{(a) Same \Tone ~experiment as shown in Fig.~\ref{fig:fakeT1}, but with DD sequence. In this experiment, we deal with the overlap of \Tone ~flip and DD pulses the same way as to measure \Ttwo, demonstrating the same RTN environment when we apply DD sequence and protect \Nit. Red diamond: experiment; black solid line: simulation; gray dashed line: fit. (b) natural \Tone ~measurement under a DD sequence with $\tau=200$ns interval (red diamond). The fit to an exponential decay (gray dashed line) gives $3.7\pm 1.3$ms in good agreement with the \Tone~ measured in the absence of DD pulses. All error bars are one SEM.}
\end{figure*}

In order to simulate a \Tone ~flip, we apply a $\pi$ pulse, the same pulse used for DD. As the pulse length ($44$ns) is comparable to the smallest time interval between pulses  of the DD sequence ($200$ns), there is a non-negligible probability that a \Tone ~flip overlaps with DD pulses for some of the \Tone ~traces. We deal with this possibility in two ways: if the overlap of the \Tone ~flip and DD $\pi$-pulse is larger than half the pulse duration, we do not apply either pulses; also, we discard all traces that contain an overlap of less than half of the pulse duration. We verify that this strategy does not bias the overall engineered noise by measuring  \Tone ~with and without DD sequence. The results in Fig.~\ref{fig:fakeT1DD} show that the fitted \Tone $_{DD}=10.0\pm 0.4\mu$s ~is the same as \Tone ~without DD ($10.0\pm 0.4\mu$s) within the errorbars. This verifies that our treatment of overlapping $\pi$-pulses does not change the underlying physics, and that the protection of nuclear spin coherence derives from DD, rather than from changes in the engineered \Tone ~under DD.

\subsection{Natural \Tone~with DD}\label{sec:naturalwDD}
To show that even for the natural \Tone ~noise our DD method can efficiently decouple the nuclear spin from its RTN, we measure the natural \Tone ~while applying a DD sequence (Fig.~\ref{fig:naturalT1DD}). In order to have a sequence robust against flip angle error and off-resonance pulses, we employ the KDD pulse sequence~\cite{Souza11KDD},  $$KDD_\phi=\tau/2-(\pi)_{\pi/6+\phi}-\tau-(\pi)_\phi-\tau-(\pi)_{\pi/2+\phi}-\tau-(\pi)_\phi-\tau-(\pi)_{\pi/6+\phi},$$ and concatenate this $5$-pulse block following the XY-$16$ phase cycle~\cite{Gullion90}: $KDD_x-KDD_y-KDD_x-KDD_y-KDD_y-KDD_x-KDD_y-KDD_x-KDD_{\bar x}-KDD_{\bar y}-KDD_{\bar x}-KDD_{\bar y}-KDD_{\bar y}-KDD_{\bar x}-KDD_{\bar y}-KDD_{\bar x}$.
We measure up to $8$ms, corresponding to $4\times10^4$ pulses, longer than \Ttwostar. We fit the data to an exponential decay and find the $1/e$ time to be $3.7\pm 1.3$ms, matching \Tone $=4.3\pm 0.3$ms. At this $\tau$, we expect more than a factor of $3$ gain in \Ttwo, therefore there should be a net improvement in nuclear spin coherence. 
This suggests that for up to about $\sim10^5$ pulses, pulse errors in the DD do not accumulate so significantly to counteract potential gains in \Ttwo. We expect our method to  protect \Nit ~beyond the limit set by natural \Tone ~of a few ms, as it could be verified with single-shot readout of the nuclear spin coherence.

\subsection{Discussion on the relation between \Tone~ and \Ttwostar}
Given the large uncertainty in the measured natural \Tone $=4.3(3)$ms and \Ttwostar~$=5.6(1.7)$ms (errors are $95\%$ confidence interval from fit), we find a   \Ttwostar $/$\Tone ~ratio of $5.6/4.3=1.3(2)$, which is compatible with the 3LF model prediction (\Ttwostar $/$\Tone $=1.5$), but does not exclude other models. 
It is then worth to examine more in depth whether the data does indeed support the  three-level spin-fluctuator model  or whether other models could be a better match.

Instead of extracting \Tone~and \Ttwostar~ independently from the fit, we fit the two experimental datasets together to four models: 1) fixing \Ttwostar $/$\Tone $=1.5$; 2) fixing \Ttwostar $/$\Tone $=2$; 3) fixing \Ttwostar $/$\Tone $=1$; 4) leaving \Ttwostar $/$\Tone~as a free fitting parameter. We compare the mean square error (MSE) of the fit and the uncertainty of the fitted \Tone ~for the four models; results are summarized  in Table.~\ref{tab:comparemodels}. This analysis reveals that the  model assuming \Ttwostar $/$\Tone $=1.5$ (model 1) yields the best result both in terms of smallest MSE and smallest uncertainty in \Tone. A more general fit (model 4) converges to \Ttwostar $/$\Tone $=1.3(5)$, similar to the result when the two datasets are fitted independently, but results in larger uncertainty in \Tone. We therefore argue that the experimental data, although with a relatively large uncertainty, are consistent with the spin-fluctuator model predictions.

\begin{table}[]
\centering
\caption{Comparing different models for $T_2^{*n}/T_1^e$}
\label{tab:comparemodels}
\begin{tabular}{lcrrc}
        & MSE & $T_1^e$(ms) & $\sigma_{T_1^e}$(ms)& $T_2^{*n}/T_1^e$\\
model 1 &0.0083&4.1&0.8&3/2\\
model 2 &0.0091&3.6&0.8&2\\
model 3 &0.0086&5.0&1.0&1\\
model 4 &0.0084&4.3&1.2&1.3(5) 
\end{tabular}
\end{table}

We further note that we can ascribe the slightly smaller \Ttwostar ~than expected (\Ttwostar $/$\Tone $<1.5$)  to environmental drift. Each data point in Fig.~\ref{fig:naturalT2star} is averaged over $100,000$ repetitions. We observe $\sim20$Hz drift  in the nuclear Larmor frequency even after recalibrating the experiment about every hour by measuring the magnetic field  drift with the NV  and compensating its effect on the nuclear Larmor frequency by adding  a corresponding phase shift to the second $\pi/2$ pulse of the nuclear Ramsey sequence. Although small, the frequency drift is non-negligible compared to the detuning in the nuclear Ramsey experiment ($\sim 800$Hz). The resulting off-resonance pulses  cause the average data to have a reduced contrast at long time, which is interpreted as a shorter \Ttwostar.

\section{Appendix 2: Coherence decay for random telegraph noise}\label{sec:appendixcoherence}
\subsection{Master Equation Description of the Nuclear Coherence}\label{sec:Lindblad}
While the random walker model provides an intuitive semiclassical picture of the decay process, one could also solve the coupled fluctuator-qubit dynamics with a fully quantum mechanical framework. In particular, as the \Tone ~process of the NV center is purely Markovian, it is valid to describe the combined electron-nuclear spin dynamics using a master equation. 
We can then write a Lindblad master equation 
\begin{equation}
\frac{d\rho}{dt}=\mathcal{L}[\rho]
=-i[\mathcal{H}_i, \rho]+\sum_{k=1}^{M}(L_k\rho(t)L_k^\dagger-\frac{1}{2}L_k^\dagger L_k \rho(t)-\frac{1}{2} \rho(t) L_k^\dagger L_k),
\end{equation}
where the jump operators $L_k$ describe the \Tone ~flips of NV, and are therefore $L_k=\Gamma\ket{m_s}\bra{m_s'}$, where $m_s, m_s'=\{-1,0,+1\}$. $\Gamma=1/\sqrt{2T_1^e}$ ($1/\sqrt{3T_1^e}$) for 2LF (3LF). 
We note that we do not need to explicitly write jump operators for the nuclear spin, as its decoherence is mediated by the Hamiltonian 
$\mathcal{H}_2=\omega_e S_z + \omega_n I_z + S\cdot \mathcal{A} \cdot I$ ($\mathcal{H}_3=D S_z^2 + \omega_e S_z + \omega_n I_z + S\cdot \mathcal{A} \cdot I$)  for the electron-nuclear spin register for an electronic spin-$1/2$ (spin-$1$). The evolved density operator can be simply found by  vectorizing this equation to obtain $\rho(t)=e^{\mathcal{L}t}\rho(0)$. 
We find that the numerical results from the spin-fluctuator model and the master equation match, indicating the validity of using the semiclassical spin-fluctuator model to describe a fully quantum process. We note that the quantum mechanical treatment could handle more general cases, such as the initial NV state being a superposition state~\cite{Wold12}.

\subsection{Analytical results for the coherence time due to a 2LF}\label{sec:T22LF}
In the 2LF case, 
we can obtain an intuitive picture of the dynamics under DD by diagonalizing the block $[e^{M_{-1}\tau/2}\cdot U_\pi \cdot e^{M_{-1}\tau/2}]=V^{-1}\Lambda_{dd} V$. The diagonal elements of $\Lambda_{dd}$ are real: 
\begin{equation}\label{eq:diagblock}
\lambda^\pm=\frac{e^{-\gamma \tau} }{{W}} [\gamma \sin({W}\tau) \pm \sqrt{v^2 - \gamma^2 \cos^2({W}\tau)} ~],
\end{equation}
with $W=\sqrt{v^2-\gamma^2}$. 
They satisfy $\lambda^+\geq 0 \geq \lambda^-$, with the equal sign only for $v=\gamma$. As we do not expect a net growth of coherence,  the negative eigenvalue $\lambda^-$ is not expected to contribute. As we will see soon, its coefficient is almost $0$. For an initial state $\vec{P}=[1;p]$, $p\in [-1,1]$, we can express the coherence explicitly: 
\begin{equation}\label{eq:diagfull}
\begin{split}
\langle e^{i\phi}(N\tau)\rangle =& \{V \Lambda_{dd}^n V^{-1} \vec{P}\}_1\\
=&p c_1 \lambda_+^n  - p c_2 \lambda_-^n  + c_3 \lambda_+^n   + c_4 \lambda_-^n  
\end{split}
\end{equation}
with $c_n$ as follows
\begin{equation}\label{eq:diagcoeff}
\begin{split}
c_1=&c_2=iv\gamma \sin^2({W}\tau/2) B, \\
c_3=&\frac{1}{2}+\frac{B}{2}[v^2-\gamma^2\cos({W}\tau)],\\
c_4=&\frac{1}{2}-\frac{B}{2}[v^2-\gamma^2\cos({W}\tau)],
\end{split}
\end{equation}
where $B=1/[W\sqrt{v^2-\gamma^2\cos^2({W}\tau)}]$. 
In the strong fluctuator regime,  $v\gg\gamma$, we have $c_1, c_2\sim \gamma/v\approx 0$, $c_3\approx 1$, $c_4\approx 0$, yielding an exponential decay $\langle e^{i\phi}(N\tau)\rangle \approx \lambda_+^n$. Note that Equations (\ref{eq:diagblock}-\ref{eq:diagcoeff}) are valid even in the weak fluctuator regime.

\subsection{Analytical results for the coherence time due to a 3LF}\label{sec:T23LF}
\begin{figure*}
    \centering
    \begin{subfigure}[t]{0.48\textwidth}
        \caption{}\label{fig:decaycomp}
        \includegraphics[width=0.8\textwidth]{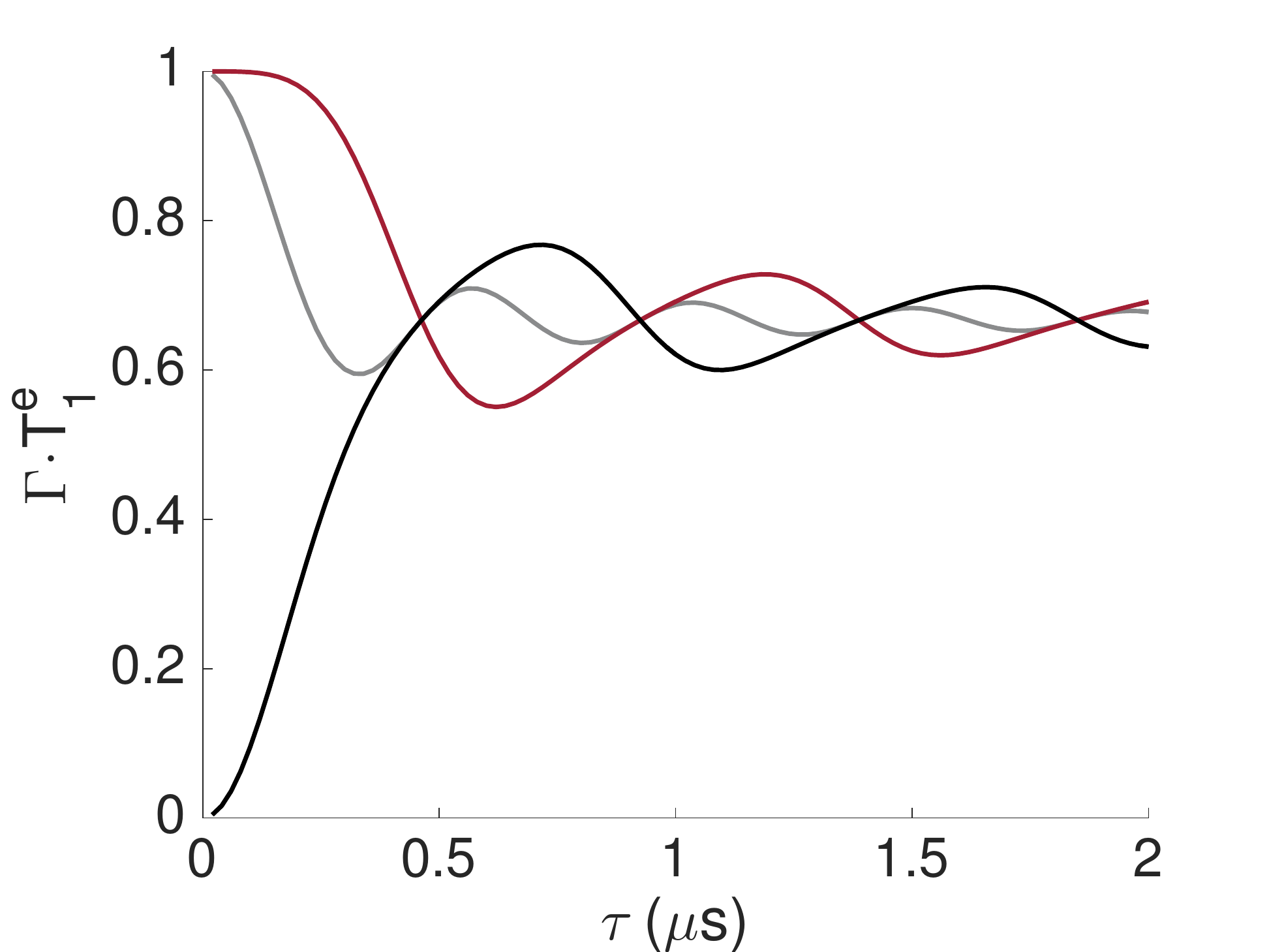}
    \end{subfigure}%
    ~ 
    \begin{subfigure}[t]{0.48\textwidth}
        \caption{}\label{contricomp}
        \includegraphics[width=0.8\textwidth]{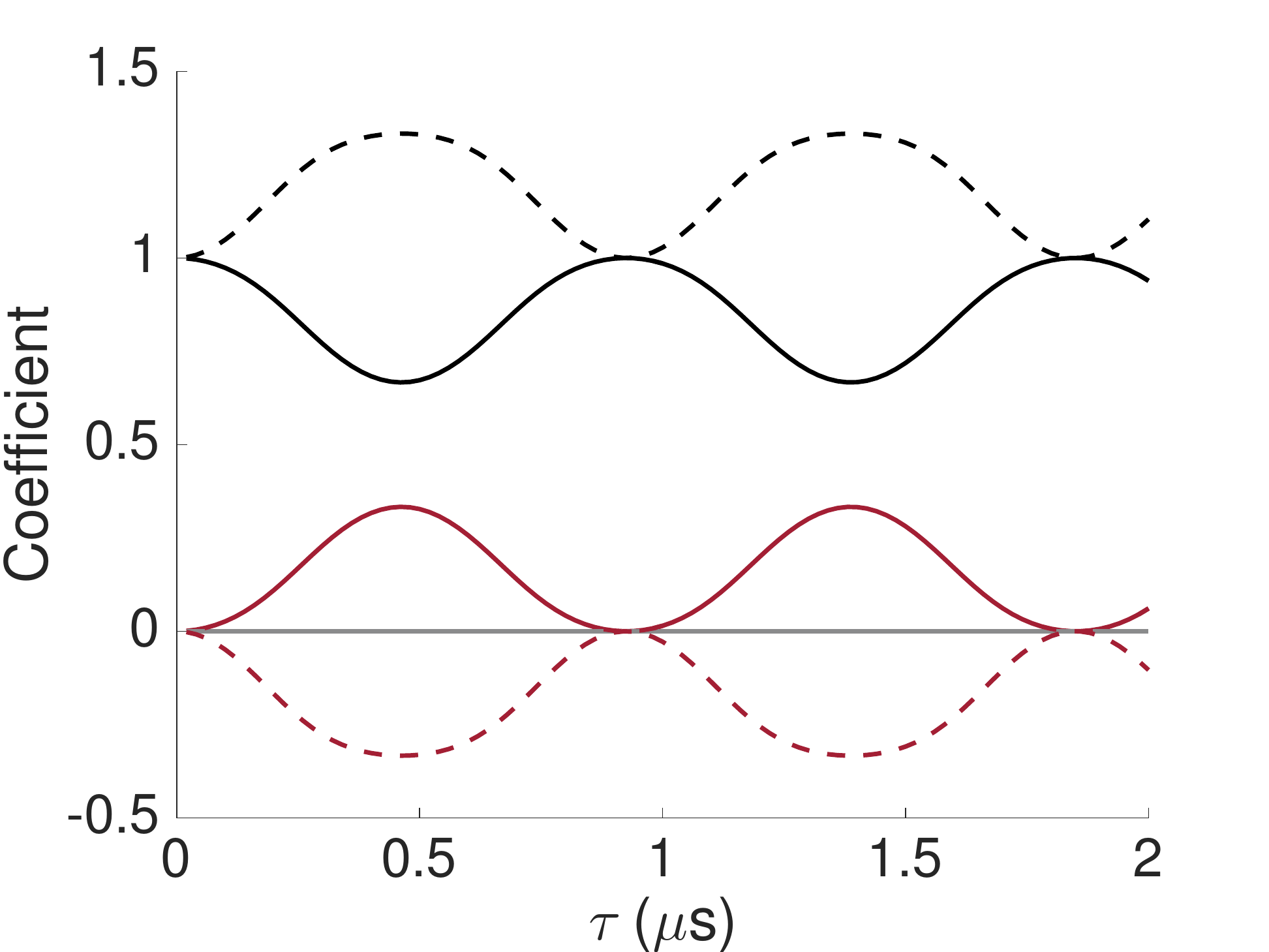}
    \end{subfigure}
    \caption{Contribution of three eigenvalues to the \Ttwo ~decay due to a 3LF. (a)  Decay rate corresponding to each eigenvalue as a function of the DD  interval $\tau$. Note the gray curve represents the negative eigenvalue, and here we plot its  absolute value. (b) Contribution of each eigenvalue to the qubit coherence. We see the negative eigenvalue has almost zero contribution. Solid lines: NV starts in the subspace spanned by $\ket{m_s=\pm 1}$; Dashed lines: NV starts in $\ket{m_s=0}$.}\label{fig:3-level_T2n}
\end{figure*}

The 3-level fluctuator case is more complicated, and we cannot derive an elegant analytical form. When we apply the DQ drive, the coherence approximately follows a simple form 
\begin{equation}\label{eq:3levelT2}
\langle e^{i\phi}(n\tau)\rangle \approx c_+ \lambda_+^n + c_0 \lambda_0^n + c_- \lambda_-^n,
\end{equation}
where $\{\lambda_+,\lambda_0,\lambda_-\}$ are the eigenvalues of $e^{M_{-1}\tau/2}\cdot U_\pi \cdot e^{M_{-1}\tau/2}$. 
Instead of writing down their cumbersome expressions, to obtain some intuition in Fig.~\ref{fig:3-level_T2n} we plot  how the eigenvalues and their corresponding coefficients change as a function of $\tau$.  Similar to the 2LF case, there is one eigenvalue, $\lambda_0$, that is negative but has a vanishing contribution to the dynamics. The coherence behavior also depends on the initial state (which sets $c_k$). Assuming \Tone$=4$ms and starting from the subspace spanned by $\ket{m_s=\pm1}$, we obtain the solid lines in Fig.~\ref{fig:3-level_T2n}; the dashed lines represent the case of starting from $\ket{m_s=0}$. Interestingly, in the latter case, the coefficients will go beyond $1$ and below $0$, causing better coherence for some $\tau$ value than the best decay rate of the three eigenvalues.

\subsection{Weak fluctuator regime for 3-level system}\label{sec:weak3LF}
\begin{figure}
    \centering
    \begin{subfigure}[t]{0.48\textwidth}
    \includegraphics[width=0.8\textwidth]{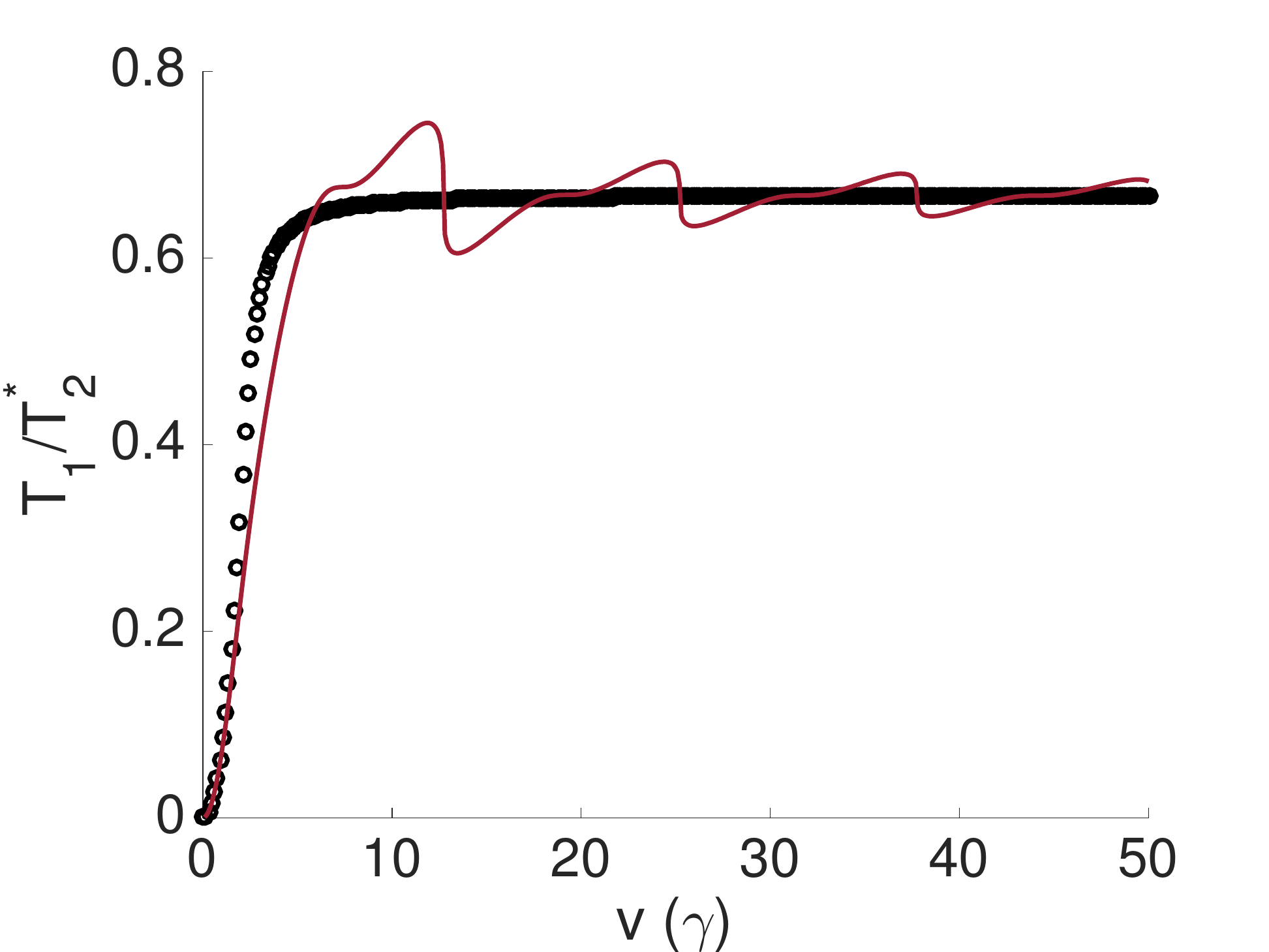}
    \end{subfigure}
    \caption{\Ttwostar ~from weak to strong fluctuator regimes as a funcion of $v/\gamma$. In the weak fluctuator regime, \Ttwostar ~increases as the fluctuator interacts more weakly with the qubit.}\label{fig:T2starweak}
\end{figure}

For completeness, we discuss briefly the 3-level weak fluctuator regime, $v\ll\gamma$ (results for the 2LF can be found in reference~\cite{Bergli07}). 
When evaluating the decay without DD (to obtain the dephasing time \Ttwostar) only one of the three eigenvalues of $M_{-1}$ contributes significantly to the decay. Therefore the decoherence has an exponential form, with 
\begin{equation}\label{eq:T2stardecaymajor}
\begin{split}
\frac{1}{T_2^*}=&2\gamma-\frac{3\gamma^2-v^2}{3^{1/3}K}-\frac{K}{3^{2/3}}\\
K=&\left(9\gamma^3+\sqrt{3}v\sqrt{v^4-9v^2\gamma^2+27\gamma^4}\right)^{1/3}
\end{split}
\end{equation}
The relationship between \Ttwostar ~and $v$ is shown in Fig.~\ref{fig:T2starweak}, covering the full range from weak fluctuator to strong fluctuator. The black circles are the approximated result from the expression in Eq.~\ref{eq:T2stardecaymajor}, which ignores small contributions from the other two eigenvalues. We note that this approximation captures the average behavior for any $v$, even if the exact result (red curve obtained from numerical calculations) shows additional features. 
Unlike in the strong fluctuator regime, where \Ttwostar$=1.5$\Tone ~is independent of $v$, \Ttwostar  ~for a weaker fluctuator  is longer in the regime where $v<\gamma$ and increases with $\gamma/v$. We also see that both in the weak and strong fluctuator regimes, an exponential decay is a good approximation, but the behavior is more complicated in the intermediate regime.

\begin{figure*}
    \centering
    \begin{subfigure}[t]{0.48\textwidth}
        \caption{}
        \includegraphics[width=0.8\textwidth]{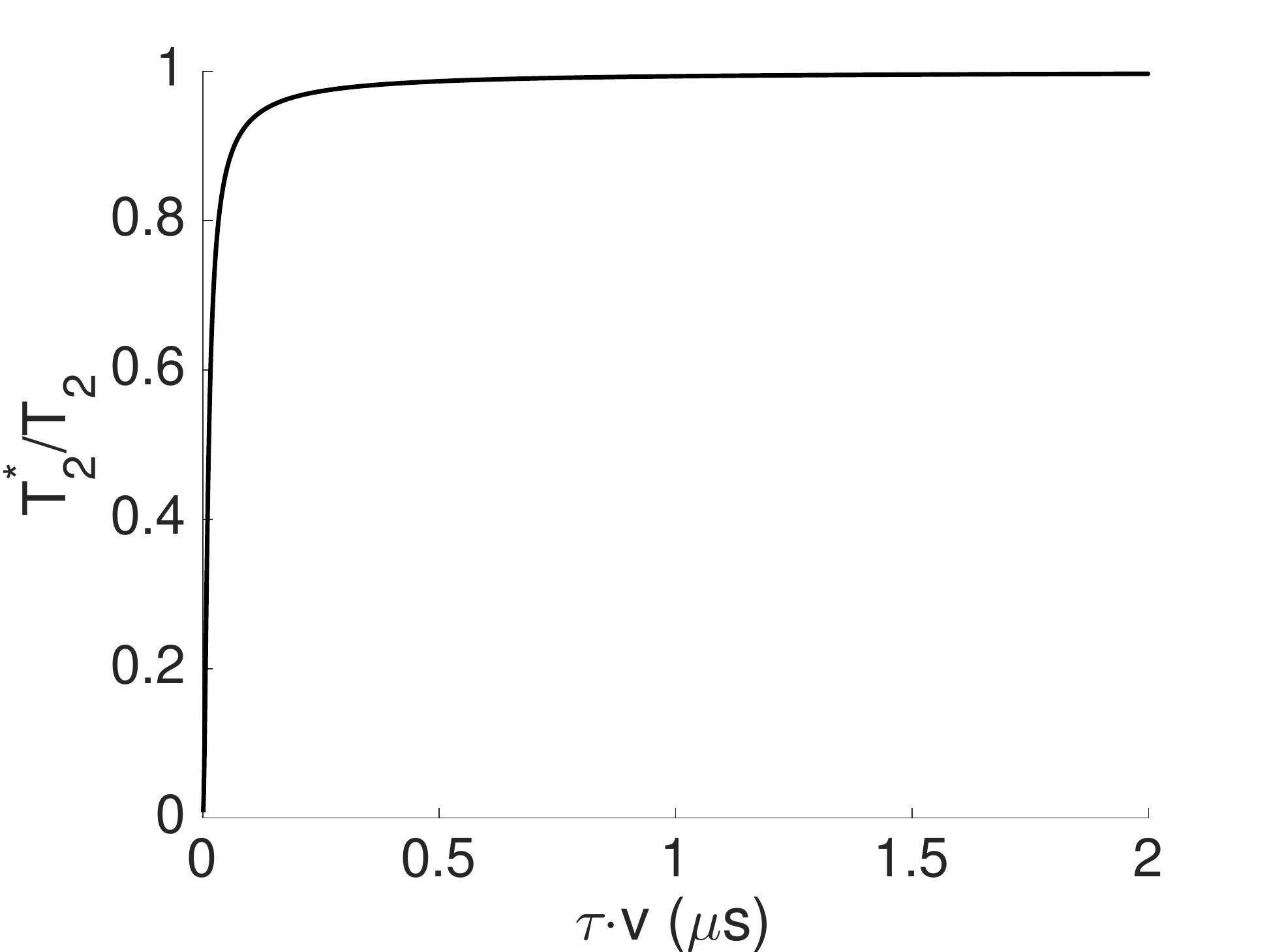}
    \end{subfigure}
    ~ 
    \begin{subfigure}[t]{0.48\textwidth}
        \caption{}
        \includegraphics[width=0.8\textwidth]{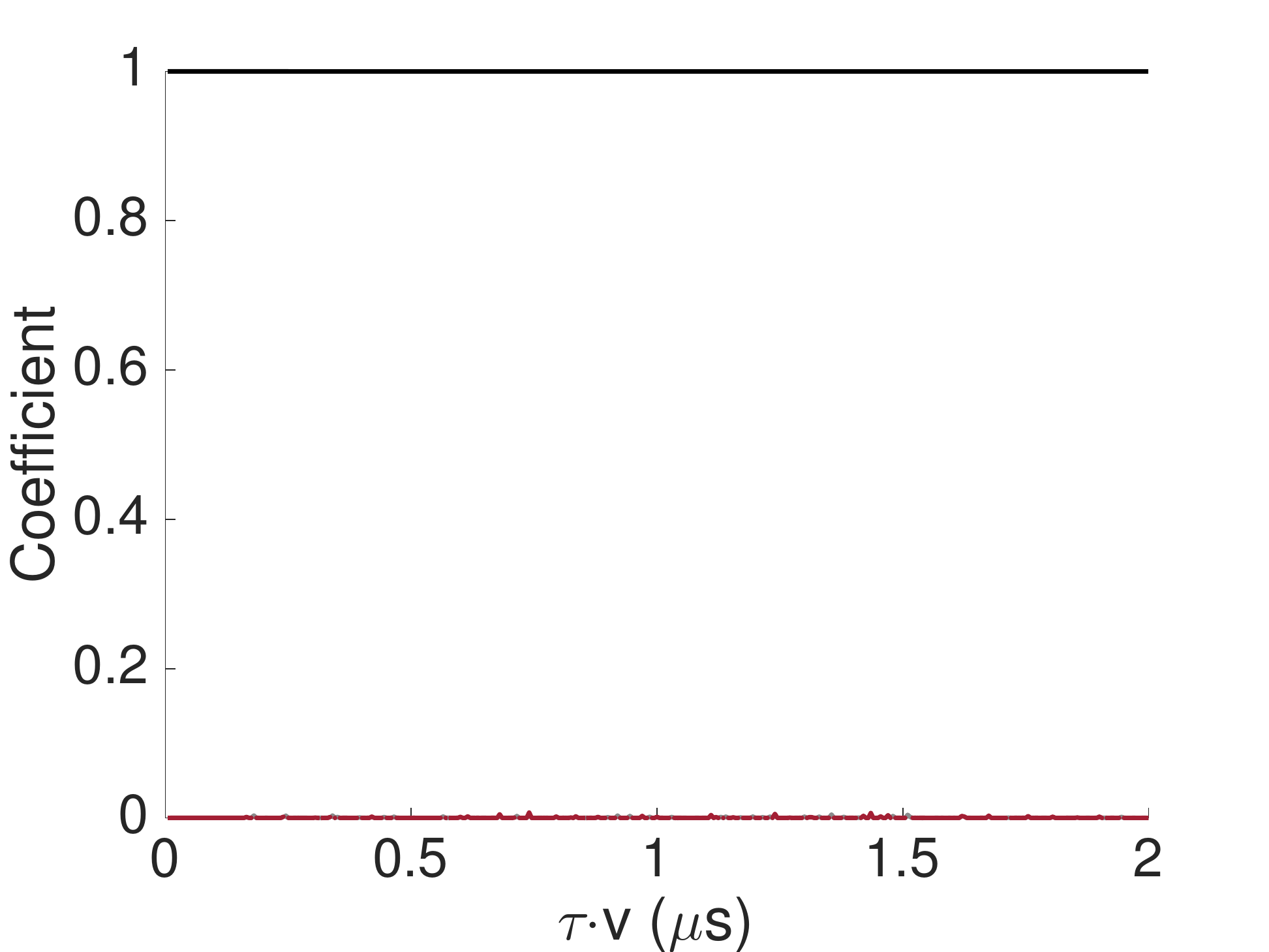}
    \end{subfigure}
    \caption{Contribution of different eigenvalues in the \Ttwo ~decay with a 3-level RTN, for weak fluctuator. (a):  decay rate corresponding to the slowest decaying eigenvalue as a function of the DD  interval $\tau$. (b): Contribution of each eigenvalue to the qubit coherence. We see that only the black line corresponding to slowest decay has nonzero contribution. Solid lines: NV start in subspace spanned by $\ket{m_s=\pm 1}$; Dashed lines (superimposed): NV starts in $\ket{m_s=0}$.}\label{fig:3-level_T2n_weak}
\end{figure*}

When we apply DD to protect the nuclear spin qubit in the weak fluctuator regime,  we find that only one eigenvalue mainly contributes to the decay, while we had two eigenvalues contributing in the strong coupling regime. The decoherence process is thus exponential. Interestingly, the main contribution  comes from the slowest decay term.  In Fig.~\ref{fig:3-level_T2n_weak}, we show example of the decay component (we did not plot the other two fast decaying components because they are many orders larger) and their contributions to the total decay under DQ drive, similar to Fig.~\ref{fig:3-level_T2n}. Here we take \Tone$=10\mu$s, $v=0.01\gamma$. A  different result due to initial state as in strong fluctuator regime is not seen here (solid lines overlap with dashed lines). The oscillatory behavior seen in Fig.~\ref{fig:DDsimulation} and Fig.~\ref{fig:decaycomp} is also missing for the weak fluctuator.

\subsection{\Ttwostar~ in the strong fluctuator regime}\label{sec:strongfluctuatorregime}
In the strong fluctuator regime, \Ttwostar~ becomes independent of $v$ (the fluctuator/qubit coupling). This had been observed in Ref.~\cite{Bergli07} for the TLF. For the 3LF, we can find a similar result 
by taking the limit of $v\gg\gamma$ in Eq.~\ref{eq:T2stardecaymajor}, 
\begin{equation}
\begin{split}
\lim_{v\gg\gamma}\frac{1}{T_2^*}&=2\gamma=\frac{2}{3T_1^e}\\
T_2^{*n}&=\frac{3}{2}T_1^e\quad{}\text{(3LF)}
\end{split}
\end{equation} 
Similarly, for 2LF, the two eigenvalues of $M_{-1}$ are $\gamma\pm\sqrt{\gamma^2-v^2}$. In the strong fluctuator regime, the only real contribution to decay is $\gamma$, 
\begin{equation}
T_2^{*n}=\frac{1}{\gamma}=2T_1^e\quad{}\text{(2LF)}.
\end{equation}

\section*{References}
\bibliographystyle{unsrt}
\bibliography{Biblio}

\begin{thebibliography}{10}

\bibitem{Devoret13}
M.~H. Devoret and R.~J. Schoelkopf.
\newblock Superconducting circuits for quantum information: An outlook.
\newblock {\em Science}, 339(6124):1169--1174, 2013.

\bibitem{Doherty13}
Marcus~W. Doherty, Neil~B. Manson, Paul Delaney, Fedor Jelezko, Joerg
  Wrachtrup, and Lloyd~C.L. Hollenberg.
\newblock The nitrogen-vacancy colour centre in diamond.
\newblock {\em Physics Reports}, 528(1):1 -- 45, 2013.

\bibitem{Degen17}
C.~L. Degen, F.~Reinhard, and P.~Cappellaro.
\newblock Quantum sensing.
\newblock {\em Rev. Mod. Phys.}, 89:035002, Jul 2017.

\bibitem{Zwanenburg13}
Floris~A. Zwanenburg, Andrew~S. Dzurak, Andrea Morello, Michelle~Y. Simmons,
  Lloyd C.~L. Hollenberg, Gerhard Klimeck, Sven Rogge, Susan~N. Coppersmith,
  and Mark~A. Eriksson.
\newblock Silicon quantum electronics.
\newblock {\em Rev. Mod. Phys.}, 85:961--1019, Jul 2013.

\bibitem{Bernien17}
H.~{Bernien}, S.~{Schwartz}, A.~{Keesling}, H.~{Levine}, A.~{Omran},
  H.~{Pichler}, S.~{Choi}, A.~S. {Zibrov}, M.~{Endres}, M.~{Greiner},
  V.~{Vuleti{\'c}}, and M.~D. {Lukin}.
\newblock Probing many-body dynamics on a 51-atom quantum simulator.
\newblock {\em Nature}, 551:579--584, November 2017.

\bibitem{Terhal15}
Barbara~M. Terhal.
\newblock Quantum error correction for quantum memories.
\newblock {\em Rev. Mod. Phys.}, 87:307--346, Apr 2015.

\bibitem{Waldherr14}
G.~Waldherr, Y.~Wang, S.~Zaiser, M.~Jamali, T.~Schulte-Herbruggen, H.~Abe,
  T.~Ohshima, J.~Isoya, J.~F. Du, P.~Neumann, and J.~Wrachtrup.
\newblock Quantum error correction in a solid-state hybrid spin register.
\newblock {\em Nature}, 506(7487):204--207, February 2014.

\bibitem{Taminiau14}
H.T. Taminiau, J.~Cramer, T.~van~der Sar, V.V. Dobrovitski, and R.~Hanson.
\newblock Universal control and error correction in multi-qubit spin registers
  in diamond.
\newblock {\em Nat Nano}, 9(3):171--176, March 2014.

\bibitem{Cramer16}
J.~Cramer, N.~Kalb, M.~A. Rol, B.~Hensen, M.~S. Blok, M.~Markham, D.~J.
  Twitchen, R.~Hanson, and T.~H. Taminiau.
\newblock Repeated quantum error correction on a continuously encoded qubit by
  real-time feedback.
\newblock {\em Nature Communications}, 7:11526, may 2016.

\bibitem{Linke17}
Norbert~M. Linke, Mauricio Gutierrez, Kevin~A. Landsman, Caroline Figgatt,
  Shantanu Debnath, Kenneth~R. Brown, and Christopher Monroe.
\newblock Fault-tolerant quantum error detection.
\newblock {\em Science Advances}, 3(10), 2017.

\bibitem{Ofek16}
N.~{Ofek}, A.~{Petrenko}, R.~{Heeres}, P.~{Reinhold}, Z.~{Leghtas},
  B.~{Vlastakis}, Y.~{Liu}, L.~{Frunzio}, S.~M. {Girvin}, L.~{Jiang},
  M.~{Mirrahimi}, M.~H. {Devoret}, and R.~J. {Schoelkopf}.
\newblock Extending the lifetime of a quantum bit with error correction in
  superconducting circuits.
\newblock {\em Nature}, 536:441--445, August 2016.

\bibitem{Hahn50}
E.~L. Hahn.
\newblock Spin echoes.
\newblock {\em Phys. Rev.}, 80(4):580--594, 1950.

\bibitem{Khodjasteh09l}
Kaveh Khodjasteh and Lorenza Viola.
\newblock Dynamically error-corrected gates for universal quantum computation.
\newblock {\em Phys. Rev. Lett.}, 102:080501, Feb 2009.

\bibitem{Cappellaro09}
P.~Cappellaro, L.~Jiang, J.~S. Hodges, and M.~D. Lukin.
\newblock Coherence and control of quantum registers based on electronic spin
  in a nuclear spin bath.
\newblock {\em Phys. Rev. Lett.}, 102(21):210502, 2009.

\bibitem{Zhang14c}
Jingfu Zhang, Alexandre~M. Souza, Frederico~Dias Brandao, and Dieter Suter.
\newblock Protected quantum computing: Interleaving gate operations with
  dynamical decoupling sequences.
\newblock {\em Phys. Rev. Lett.}, 112:050502, Feb 2014.

\bibitem{Paz-Silva13}
Gerardo~A Paz-Silva and D~A Lidar.
\newblock Optimally combining dynamical decoupling and quantum error
  correction.
\newblock {\em Scientific Reports}, 3:1530, apr 2013.

\bibitem{Byrd02}
Mark~S. Byrd and Daniel~A. Lidar.
\newblock Comprehensive encoding and decoupling solution to problems of
  decoherence and design in solid-state quantum computing.
\newblock {\em Phys. Rev. Lett.}, 89:047901, Jul 2002.

\bibitem{Boulant02}
N.~Boulant, M.~A. Pravia, E.~M. Fortunato, T.~F. Havel, and D.~G. Cory.
\newblock Experimental concatenation of quantum error correction with
  decoupling.
\newblock {\em Quantum Information Processing}, 1(1):135--144, Apr 2002.

\bibitem{Biercuk11}
M~J Biercuk, A~C Doherty, and H~Uys.
\newblock Dynamical decoupling sequence construction as a filter-design
  problem.
\newblock {\em J. of Phys. B}, 44(15):154002, 2011.

\bibitem{Ban09}
Masashi Ban, Sachiko Kitajima, and Fumiaki Shibata.
\newblock Dynamical suppression of dephasing for {Markov} processes.
\newblock {\em Phys. Lett. A}, 373(40):3614--3618, sep 2009.

\bibitem{Paladino14}
E.~Paladino, Y.~M. Galperin, G.~Falci, and B.~L. Altshuler.
\newblock 1/f noise: Implications for solid-state quantum information.
\newblock {\em Rev. Mod. Phys.}, 86:361--418, Apr 2014.

\bibitem{Bermeister14}
Adam Bermeister, Daniel Keith, and Dimitrie Culcer.
\newblock Charge noise, spin-orbit coupling, and dephasing of single-spin
  qubits.
\newblock {\em App. Phys. Lett}, 105(19):192102, 2014.

\bibitem{Burnett14}
J.~Burnett, L.~Faoro, I.~Wisby, V.~L. Gurtovoi, A.~V. Chernykh, G.~M.
  Mikhailov, V.~A. Tulin, R.~Shaikhaidarov, V.~Antonov, P.~J. Meeson, A.~Ya.
  Tzalenchuk, and T.~Lindstroem.
\newblock Evidence for interacting two-level systems from the 1/f noise of a
  superconducting resonator.
\newblock {\em Nat. Commun.}, 5:4119--, June 2014.

\bibitem{deGraaf17}
S.~E. de~Graaf, A.~A. Adamyan, T.~Lindstr\"om, D.~Erts, S.~E. Kubatkin, A.~Ya.
  Tzalenchuk, and A.~V. Danilov.
\newblock Direct identification of dilute surface spins on
  ${\mathrm{al}}_{2}{\mathrm{o}}_{3}$: Origin of flux noise in quantum
  circuits.
\newblock {\em Phys. Rev. Lett.}, 118:057703, Jan 2017.

\bibitem{Maurer12}
P.~C. Maurer, G.~Kucsko, C.~Latta, L.~Jiang, N.~Y. Yao, S.~D. Bennett,
  F.~Pastawski, D.~Hunger, N.~Chisholm, M.~Markham, D.~J. Twitchen, J.~I.
  Cirac, and M.~D. Lukin.
\newblock Room-temperature quantum bit memory exceeding one second.
\newblock {\em Science}, 336(6086):1283--1286, 2012.

\bibitem{Dolde14}
Florian Dolde, Ville Bergholm, Ya~Wang, Ingmar Jakobi, Boris Naydenov,
  Sebastien Pezzagna, Jan Meijer, Fedor Jelezko, Philipp Neumann, Thomas
  Schulte-Herbruggen, Jacob Biamonte, and Joerg Wrachtrup.
\newblock High-fidelity spin entanglement using optimal control.
\newblock {\em Nat Commun}, 5:--, February 2014.

\bibitem{Hirose16}
Masashi Hirose and Paola Cappellaro.
\newblock Coherent feedback control of a single qubit in diamond.
\newblock {\em Nature}, 532(7597):77--80, April 2016.

\bibitem{Aslam17}
Nabeel Aslam, Matthias Pfender, Philipp Neumann, Rolf Reuter, Andrea Zappe,
  Felipe F{\'a}varo~de Oliveira, Andrej Denisenko, Hitoshi Sumiya, Shinobu
  Onoda, Junichi Isoya, and J{\"o}rg Wrachtrup.
\newblock Nanoscale nuclear magnetic resonance with chemical resolution.
\newblock {\em Science}, page eaam8697, 2017.

\bibitem{Zaiser16}
Sebastian Zaiser, Torsten Rendler, Ingmar Jakobi, Thomas Wolf, Sang-Yun Lee,
  Samuel Wagner, Ville Bergholm, Thomas Schulte-Herbruggen, Philipp Neumann,
  and Jorg Wrachtrup.
\newblock Enhancing quantum sensing sensitivity by a quantum memory.
\newblock {\em Nat. Commun.}, 7:--, August 2016.

\bibitem{Shim13}
J.~H. Shim, I.~Niemeyer, J.~Zhang, and D.~Suter.
\newblock Room-temperature high-speed nuclear-spin quantum memory in diamond.
\newblock {\em Phys. Rev. A}, 87:012301, Jan 2013.

\bibitem{Jiang08}
L.~Jiang, M.~V.~Gurudev Dutt, E.~Togan, L.~Childress, P.~Cappellaro, J.~M.
  Taylor, and M.~D. Lukin.
\newblock Coherence of an optically illuminated single nuclear spin qubit.
\newblock {\em Phys. Rev. Lett.}, 100(7):073001, 2008.

\bibitem{Reiserer16}
Andreas Reiserer, Norbert Kalb, Machiel~S. Blok, Koen J.~M. van Bemmelen,
  Tim~H. Taminiau, Ronald Hanson, Daniel~J. Twitchen, and Matthew Markham.
\newblock Robust quantum-network memory using decoherence-protected subspaces
  of nuclear spins.
\newblock {\em Phys. Rev. X}, 6:021040, Jun 2016.

\bibitem{WangFan17}
F.~Wang, Y.-Y. Huang, Z.-Y. Zhang, C.~Zu, P.-Y. Hou, X.-X. Yuan, W.-B. Wang,
  W.-G. Zhang, L.~He, X.-Y. Chang, and L.-M. Duan.
\newblock Room-temperature storage of quantum entanglement using
  decoherence-free subspace in a solid-state spin system.
\newblock {\em Phys. Rev. B}, 96:134314, Oct 2017.

\bibitem{Chen15}
Mo~Chen, Masashi Hirose, and Paola Cappellaro.
\newblock Measurement of transverse hyperfine interaction by forbidden
  transitions.
\newblock {\em Phys. Rev. B}, 92:020101, Jul 2015.

\bibitem{Sangtawesin15}
S.~Sangtawesin and J.~R. Petta.
\newblock Hyperfine-enhanced gyromagnetic ratio of a nuclear spin in diamond,
  2015.

\bibitem{Neumann10b}
Philipp Neumann, Johannes Beck, Matthias Steiner, Florian Rempp, Helmut Fedder,
  Philip~R. Hemmer, Jorg Wrachtrup, and Fedor Jelezko.
\newblock Single-shot readout of a single nuclear spin.
\newblock {\em Science}, 5991:542--544, 2010.

\bibitem{Bergli07}
Joakim Bergli and Lara Faoro.
\newblock Exact solution for the dynamical decoupling of a qubit with telegraph
  noise.
\newblock {\em Phys. Rev. B}, 75(5), feb 2007.

\bibitem{Jacques09}
V.~Jacques, P.~Neumann, J.~Beck, M.~Markham, D.~Twitchen, J.~Meijer, F.~Kaiser,
  G.~Balasubramanian, F.~Jelezko, and J.~Wrachtrup.
\newblock Dynamic polarization of single nuclear spins by optical pumping of
  nitrogen-vacancy color centers in diamond at room temperature.
\newblock {\em Phys. Rev. Lett.}, 102(5):057403, 2009.

\bibitem{Meiboom58}
S.~Meiboom and D.~Gill.
\newblock Modified spin-echo method for measuring nuclear relaxation times.
\newblock {\em Rev. Sc. Instr.}, 29(8):688--691, 1958.

\bibitem{Fuchs09}
G.~D. Fuchs, V.~V. Dobrovitski, D.~M. Toyli, F.~J. Heremans, and D.~D.
  Awschalom.
\newblock Gigahertz dynamics of a strongly driven single quantum spin.
\newblock {\em Science}, 326(5959):1520--1522, 2009.

\bibitem{DeLange10}
G.~de~Lange, Z.~H. Wang, D.~Riste, V.~V. Dobrovitski, and R.~Hanson.
\newblock Universal dynamical decoupling of a single solid-state spin from a
  spin bath.
\newblock {\em Science}, 330(6000):60--3, October 2010.

\bibitem{Scheuer14}
Jochen Scheuer, Xi~Kong, Ressa~S Said, Jeson Chen, Andrea Kurz, Luca Marseglia,
  Jiangfeng Du, Philip~R Hemmer, Simone Montangero, Tommaso Calarco, Boris
  Naydenov, and Fedor Jelezko.
\newblock Precise qubit control beyond the rotating wave approximation.
\newblock {\em New J. Phys.}, 16(9):093022, 2014.

\bibitem{Souza11KDD}
Alexandre~M. Souza, Gonzalo~A. \'Alvarez, and Dieter Suter.
\newblock Robust dynamical decoupling for quantum computing and quantum memory.
\newblock {\em Phys. Rev. Lett.}, 106:240501, Jun 2011.

\bibitem{Robledo11}
Lucio Robledo, Lilian Childress, Hannes Bernien, Bas Hensen, Paul F.~A.
  Alkemade, and Ronald Hanson.
\newblock High-fidelity projective read-out of a solid-state spin quantum
  register.
\newblock {\em Nature}, 477(7366):574--578, 2011.

\bibitem{Cohen17}
I.~{Cohen}, T.~{Unden}, F.~{Jelezko}, and A.~{Retzker}.
\newblock Protecting a nuclear spin from a noisy electron spin in diamond.
\newblock {\em ArXiv: 1703.01596}, March 2017.

\bibitem{simmonds04}
R.W. Simmonds, K.~M. Lang, D.~A. Hite, S.~Nam, D.~P. Pappas, and John~M.
  Martinis.
\newblock Decoherence in josephson phase qubits from junction resonators.
\newblock {\em Phys. Rev. Lett.}, 93:077003, 2004.

\bibitem{Gullion90}
Terry Gullion, David~B Baker, and Mark~S Conradi.
\newblock New, compensated carr-purcell sequences.
\newblock {\em J. Mag. Res.}, 89(3):479 -- 484, 1990.

\bibitem{Wold12}
Henry~J. Wold, H\aa{}kon Brox, Yuri~M. Galperin, and Joakim Bergli.
\newblock Decoherence of a qubit due to either a quantum fluctuator, or
  classical telegraph noise.
\newblock {\em Phys. Rev. B}, 86:205404, Nov 2012.

\end{thebibliography}

\end{document}